\documentclass[APA,Times1COL]{WileyNJDv5}

\usepackage{tcolorbox}
\usepackage{float}

\articletype{Research Article}%

\received{Date Month Year}
\revised{Date Month Year}
\accepted{Date Month Year}
\journal{Journal of Engineering Education}
\volume{00}
\copyyear{2025}
\startpage{1}

\begin{document}

\title{Methodological Foundations for AI-Driven Survey Question Generation}

\author[1]{Ted K. Mburu}

\author[2]{Kangxuan Rong}

\author[2]{Campbell J. McColley}

\author[2]{Alexandra Werth}

\authormark{MBURU \textsc{et al.}}
\titlemark{Methodological Foundations for AI-Driven Survey Question Generation}

\address[1]{\orgdiv{Engineering Education Program}, \orgname{University of Colorado Boulder}, \orgaddress{\state{Colorado}, \country{United States of America}}}

\address[2]{\orgdiv{Meinig School of Biomedical Engineering}, \orgname{Cornell University}, \orgaddress{\state{New York}, \country{United States of America}}}

\corres{Alexandra Werth, Meinig School of Biomedical Engineering, Cornell University, Ithaca, NY, 14850, United States of America \email{alexandra.werth@cornell.edu}}

\fundingInfo{Cornell Center for Teaching Innovation, Innovative Teaching \& Learning Grant, Cornell University}

\abstract[Abstract]{\noindent\textbf{Background:} This study investigates the use of Large Language Models to create adaptive, contextually relevant survey questions, aiming to enhance data quality in educational research without limiting scalability.

\noindent \textbf{Purpose:} We provide step-by-step methods to develop a dynamic AI-driven survey instrument and introduce the Synthetic Question-Response Analysis (SQRA) framework, a methodology designed to help evaluate AI-generated questions before deployment with human participants. 

\noindent \textbf{Design:} We examine the questions generated by our survey instrument, as well as compare AI-to-AI, generated through our SQRA framework, with AI-to-human interactions. Activity Theory provides a theoretical lens to examine the dynamic interactions between AI and participants, highlighting the mutual influence within the survey tool.
 
\noindent \textbf{Results:} We found that AI-generated questions were contextually relevant and adaptable, successfully incorporating course-specific references. However, issues such as redundant phrasing, double-barreled questions, and jargon affected the clarity of the question. Although the SQRA framework exhibited limitations in replicating human response variability, its iterative refinement process proved effective in improving question quality, reinforcing the utility of this approach for enhancing AI-driven survey.

\noindent \textbf{Conclusions:} While AI-driven question generation can enhance the scalability and personalization of open-ended survey prompts, more research is needed to establish best practices for high-quality educational research. The SQRA framework demonstrated practical utility for prompt refinement and initial validation of AI-generated survey content, but it is not capable of replicating human responses. We highlight the importance of iterative prompt engineering, ethical considerations, and the need for methodological advancements in the development of trustworthy AI-driven survey instruments for educational research.}


\jnlcitation{\cname{%
\author{Mburu, T. K.},
\author{Rong, K.},
\author{McColley, C. J.}, and
\author{Werth, A}}.
\year{(2025)}
\ctitle{Methodological Foundations for AI-Driven Survey Question Generation} \cjournal{\it J Eng. Educ.} \cvol{ }.}

\maketitle

\renewcommand\thefootnote{}
\footnotetext{\textbf{Abbreviations:} GenAI, generative artificial intelligence; LLMs, large language models; SQRA, synthetic question-response analysis; AT, activity theory; GDPR, General Data Protection Regulation; FERPA, Family Educational Rights and Privacy Act; PII, personally identifiable information; SSRL, socially shared regulation of learning; API, application programming interface; IRB, Institutional Review Board; NLTK, Natural Language Toolkit; NLP, natural language processing}

\renewcommand\thefootnote{\fnsymbol{footnote}}
\setcounter{footnote}{1}

\section{Introduction}\label{sec1}

Generative AI (GenAI) is rapidly reshaping the landscape of educational research, offering new ways to scale, personalize, and adapt research instruments like surveys. In particular, GenAI enables dynamic question generation, potentially transforming how researchers gather and respond to participant input in real time. At the same time, its use introduces complex challenges related to bias, data privacy, and reproducibility \cite{Watkins, Kusters, lu2025machinelearningsyntheticdata, Hosseini}. These concerns must be carefully addressed to ensure that AI-driven methods support responsible, meaningful research without displacing the value of human interpretation.

\begin{figure}[h]
\centering
\includegraphics[width=0.7\textwidth]{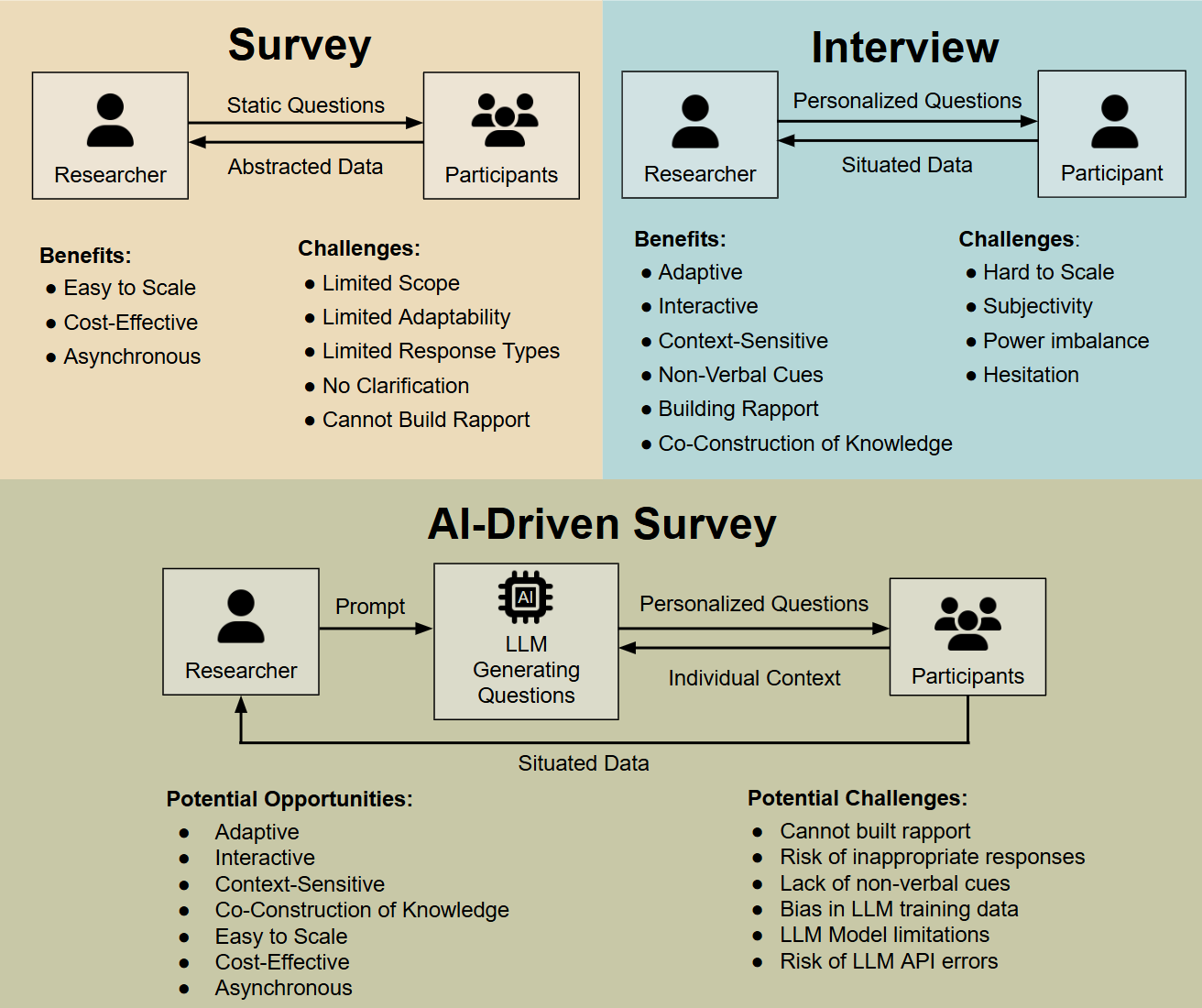}
\caption{Comparison of traditional surveys, interviews, and AI-driven surveys highlighting their benefits, challenges, and opportunities.}
\label{fig:intro_figure}
\end{figure}

This paper explores how GenAI can enhance question generation in educational survey instruments by integrating LLMs into survey design for dynamic, personalized question creation. This is especially relevant in engineering education, where the reliability and validity of survey instruments are central to high-quality research. We propose that responsive, AI-generated surveys may occupy a middle ground between traditional interviews and open-ended surveys, combining elements of both. Interviews provide adaptable, in-depth interaction but are difficult to scale; surveys are scalable and structured but lack responsiveness. AI-driven surveys aim to bridge these approaches by enabling tailored, evolving interactions while preserving efficiency. Figure \ref{fig:intro_figure} illustrates the affordances and trade-offs of each method.

\vspace{-2mm}

\vspace{-5mm}
These AI-generated surveys are well-suited to action research contexts, which emphasize iterative reflection and practice-based inquiry. Their adaptability may support more context-specific, responsive engagement, enabling researchers to tailor prompts in real time to participant input \cite{Sammut, LincolnGuba}. In these settings, well-designed AI tools can promote scalable, yet personalized, data collection.

Despite these benefits, limitations remain. One key challenge is the absence of interpersonal rappor---often central to effective interviews---which can affect participant trust and engagement \cite{Horsfall}. Without nonverbal cues or dynamic clarification, the richness of interaction may be reduced. Further, AI-generated prompts can vary in tone, relevance, or clarity, especially when poorly designed \cite{Johri, Menekse, Hosseini, Miao}. These limitations reinforce the importance of prompt design and of viewing AI surveys as complementary to---not replacements for---traditional approaches.

As the use of GenAI in research grows, so too does the need for frameworks that support transparency, validation, and trustworthiness. AI-generated questions may appear confident yet lack clear justification for why they were produced. This can be problematic in sensitive educational contexts, where transparency and alignment with learning goals are essential \cite{Kusters}. We argue that new validation procedures should be applied before such tools are deployed with students.

To that end, we introduce the Synthetic Question-Response Analysis (SQRA) framework---an AI-to-AI evaluation process where one agent generates questions and another simulates a student participant. This strategy allows researchers to analyze question quality, tone, and structure before the survey is administered to humans. We compare these simulated results with those from actual AI-to-human interactions and use sentiment, grammatical, and structural analyses to assess the framework's effectiveness.

We also apply Activity Theory (AT) to examine how the type of participant---human or simulated---shapes the survey interaction. AT provides a lens for understanding how tools mediate learning and inquiry, emphasizing the co-evolving relationship between users, tools, and context \cite{Russell, DeCosta, Georg, ilishkina2025rethinking, Hite}. This framework helps us explore how dynamic survey tools operate within broader educational systems.

Our study is guided by the following research questions:

\begin{enumerate} \item[{\bf RQ1:}] What are the key technical and ethical considerations for developing AI-driven survey systems that use Large Language Models (LLMs) to generate adaptive, personalized questions at scale? \item[{\bf RQ2:}] How do the sentiment, lexical, and grammatical characteristics of questions generated \emph{in silico} differ from those generated through human-to-AI interactions, and what does this reveal about the effectiveness of the Synthetic Question-Response Analysis (SQRA) framework? \end{enumerate}

By addressing these questions, this paper offers practical guidance for researchers interested in integrating GenAI into educational survey design, while also contributing a methodological approach for evaluating adaptive AI systems through simulation. In doing so, we engage current conversations around ethical AI use, transparency, and methodologies in education research.

\section{Background}\label{sec2}

\subsection{Emergence of Generative AI in Engineering Education Research}

Generative AI is increasingly being explored as a way to automate and streamline key research tasks such as question generation, personalized content delivery, and text analysis in education settings~\cite{Johri, Menekse, Alasadi, Baidoo}. These capabilities, enabled by LLMs, can reduce the time and expertise required for designing survey instruments, particularly when managing extensive datasets or participant pools~\cite{Johri, Menekse, Baidoo}.

By generating human-like text in response to prompts, LLMs can support the development of dynamic, context-aware survey questions with limited manual input~\cite{Johri, Menekse, Baidoo}. This automation can shift researchers' focus from manual content creation to interpreting results and refining study designs. Combined with relatively low API costs, GenAI tools offer an affordable option for scaling data collection in resource-constrained environments~\cite{VanCampenhout, Watkins, Menekse}.

GenAI-enabled surveys also align with principles of emergent design, as they can adapt in real time based on participant input. Rather than following a fixed question set, these tools generate personalized questions on the fly, capturing context-specific insights that may be missed with traditional methods~\cite{LincolnGuba}. This adaptability may be especially valuable in studies where participant experiences are diverse or difficult to predict in advance.

In this study, we use the Qualtrics platform to implement GenAI-driven surveys. Qualtrics, a widely used platform for survey design and data collection, supports API integration with tools like OpenAI's ChatGPT and provides features such as branching logic, embedded data, and secure compliance with educational data privacy regulations, e.g., General Data Protection Regulation (GDPR) and the Family Educational Rights and Privacy Act (FERPA). These capabilities allow GenAI to be embedded within a widely accepted platform for educational research.

As platforms, like Qualtrics, begin to integrate GenAI directly into survey design workflows, new possibilities emerge for real-time tailoring of questions based on student responses. Such personalization may increase engagement and provide richer qualitative data, especially in contexts where survey fatigue and surface-level responses are common~\cite{Sammut}. This level of personalization is particularly valuable in educational contexts, where many learning needs and experiences must be addressed. By supporting scalable, context-sensitive question generation, LLMs and GenAI offer a promising, cost-effective solution for advancing educational research methodologies without sacrificing quality \cite{Johri, Menekse}.

Beyond real-time personalization, GenAI also supports pre-deployment testing through synthetic data generation. Using simulated participants, researchers can evaluate how well question prompts align with educational goals and refine them before engaging real students~\cite{Johri, Alasadi}. This strategy addresses privacy concerns while offering a practical approach to prompt design validation. AI-to-AI interactions allow for iterative testing of both question quality and system responsiveness under varied conditions, creating a safer environment for experimentation prior to use in real classroom settings~\cite{Dorodchi}. This iterative process helps refine the prompt design and ensures that the AI generates contextually relevant and pedagogically sound questions \cite{lu2025machinelearningsyntheticdata}.

\subsubsection{Limitations and Ethical Concerns of AI-Driven Question Generation}

While AI-driven question generation offers promising opportunities for educational research, it also introduces significant limitations and ethical concerns that must be critically examined. One primary challenge is ensuring the accuracy and relevance of AI-generated content. Despite their advanced capabilities, LLMs may produce questions that are off-topic, misleading, or inappropriate for certain educational contexts, particularly when lacking domain-specific expertise. Moreover, the ``black-box'' nature of these models raises concerns about transparency and accountability~\cite{Kusters}. Researchers often cannot fully explain why an LLM generates a particular question, making it difficult to detect or correct biases that may arise from the model's training data~\cite{Miao}.

Bias is another key concern. Since LLMs are trained on large, real-world datasets that may reflect existing social inequities, they can inadvertently reproduce or amplify stereotypes---especially on topics related to identity, equity, or power~\cite{Johri, Menekse, Hosseini}. Without safeguards, these biases can affect the quality of data collected and potentially reinforce harmful patterns in educational research~\cite{Kusters}. Transparent documentation, prompt auditing, and iterative testing are important for identifying and mitigating these issues~\cite{Watkins, Kusters, Menekse}.

Data privacy is also a central issue. Educational research often involves sensitive participant information, and using GenAI tools responsibly requires strict adherence to privacy standards such as GDPR and FERPA~\cite{Menekse, lu2025machinelearningsyntheticdata}. AI systems should not store or use personally identifiable information without explicit consent, and extra caution is needed when third-party APIs---like OpenAI's---are used in the data pipeline. While platforms like Qualtrics offer built-in security features such as encryption and anonymization, researchers must ensure these protections extend to all tools and services involved.

Together, these limitations highlight the importance of human oversight in the design, validation, and use of AI-generated survey tools. Careful attention to prompt quality, bias mitigation, and privacy protections is essential---not only to ensure the ethical integrity of the research but also to maintain participant trust and the overall credibility of the findings.

\subsection{Activity Theory as a Theoretical Framework}

To examine the complexities of integrating AI-driven tools into educational research, we draw on Activity Theory (AT) as a guiding framework. Originally developed by Vygotsky, Luria, and Leontiev and later expanded by Engestr{\"o}m, AT serves as a theoretical framework for studying complex, tool-mediated interactions within structured activities \cite{Georg, DeCosta}. It emphasizes the interdependence of core elements in an activity system---Subject, Tool, Object, Rules, Community, and Division of Labor---highlighting how actions are shaped by cultural norms, institutional constraints, and evolving goals \cite{Engestrom, Vygotsky, leontiev}. This model emphasizes that the tool---in our case, the AI survey instrument---is not a neutral intermediary, but an active component that shapes and is shaped by interactions with the subject (e.g., the student or AI agent). 

\begin{figure}[h]
\centering
\includegraphics[width=0.8\textwidth]{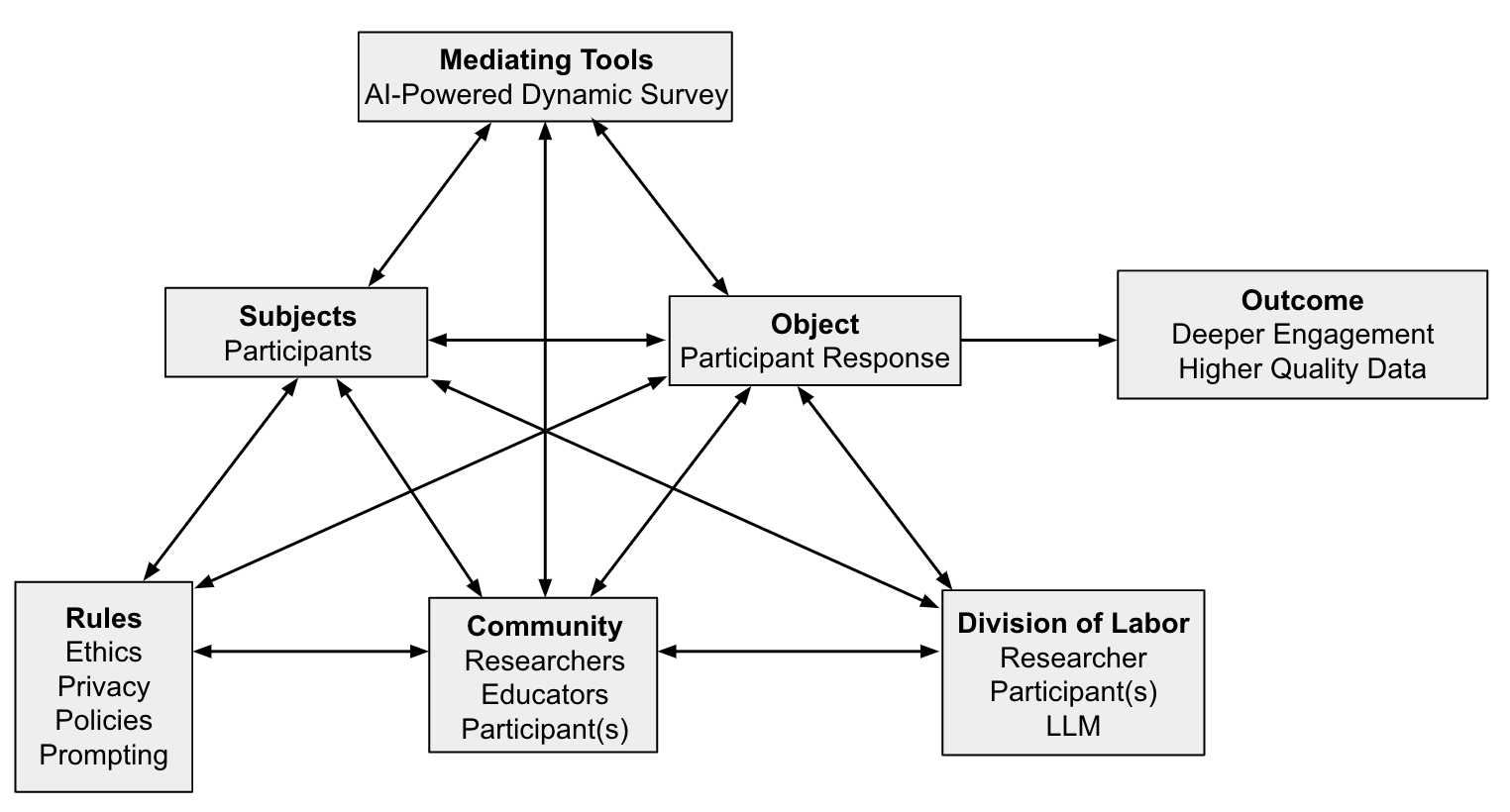}
\caption{Adaptation of Engestr{\"o}m's AT triangle for AI-driven surveys.}
\label{fig:triangle}
\end{figure}

In this study, we conceptualize the AI survey tool as a mediating artifact that not only supports but actively shapes participant interaction. The subject may be either a human student or an AI agent. The object is the student's reflective response. Rules include ethical protocols and prompt design constraints. The community includes the learning environment and norms around human-AI interaction. And the division of labor spans across researchers, student participants, and AI systems. Figure \ref{fig:triangle} illustrates these relationships using an adaptation of Engestr{"o}m’s AT triangle.

AT has been widely applied in educational and technological research to understand how tools mediate participant interactions and learning processes in technology-enhanced environments  \cite{Hite, ilishkina2025rethinking, Russell}. For instance, studies show that AT provides valuable insights into how AI-driven tools support the collaborative design and data collection processes by mediating both individual and group engagement~\cite{Georg, DeCosta}. In adaptive educational settings, AT enables researchers to examine how interactions with AI tools influence learning outcomes and data quality, helping to ensure that the tools align effectively with pedagogical objectives and research goals~\cite{ilishkina2025rethinking, Hite}. 

By systematically defining these components, researchers can use AT not only to frame their research but also to analyze the dynamic interactions between these elements. For example, AT encourages examining how the subject-tool relationship evolves under specific rules, how the community context shapes engagement, and how the division of labor influences the achievement of research goals. Its adaptability allows researchers to focus on specific elements while recognizing their interconnections.

This framework informed our prompt engineering, validation strategy, and analytical approach, enabling us to treat survey interactions not as isolated exchanges, but as embedded within broader social, institutional, and technological systems. In doing so, we join a growing body of literature that operationalizes AT in technology-enhanced learning environments.

We organize our application of AT around two core components of this study: (i) the design of AI-driven surveys and (ii) the validation of survey tools using both human and synthetic data. While our primary focus is on tool design and validation, AT also provides a useful foundation for future work analyzing how AI-mediated surveys influence participant responses and adapt over time. This structure allows AT to function as both a conceptual and practical guide for refining GenAI-driven survey instruments.

\subsubsection{AT for AI-Driven Survey Design}
In designing AI-driven surveys, the primary \textbf{subject} is the student participant. Their academic background, cognitive load, and prior experiences shape how they engage with survey prompts. These considerations influence not only question content but also delivery modality (e.g., chatbot vs. Qualtrics), tone (e.g., formal vs. conversational), and the balance between structure and adaptability (e.g., temperature settings for variability).

The \textbf{object} of the survey---student responses---must support reflection without encouraging satisficing or overly generic answers. The AI survey \textbf{tool} mediates this interaction, adapting question phrasing in response to participant input to sustain engagement and elicit relevant, thoughtful data.

\textbf{Rules}, including ethical guidelines and institutional policies, shape the interaction by requiring informed consent, protecting data privacy, and maintaining response integrity. These same constraints also inform how the AI tool is instructed---via system prompts---to generate appropriate and context-aware questions.

The \textbf{community}---including institutional norms, peer dynamics, and broader attitudes toward human-AI interaction---affects how students perceive and respond to AI-generated questions. These social factors may influence decision-making and trust in the survey design process.

The \textbf{division of labor} spans multiple roles: students provide responses, researchers design prompts and analyze data, and the AI tool facilitates question generation. Additional considerations include participant burden, the number of questions required for meaningful analysis as well as the infrastructural, human, societal, and environmental efforts needed to develop, deploy, and sustain AI systems.

\subsubsection{AT for AI-Driven Survey Validation}
Validation processes aim to ensure that AI-generated questions are meaningful, appropriate, and aligned with research goals. In this context, the \textbf{subject} includes both student participants and synthetic AI agents used in the SQRA framework. These agents help simulate and evaluate prompt performance before use in real settings.

The \textbf{object} is the student (or synthetic) response, which serves as the basis for evaluating question quality and tool adaptability. Human oversight, synthetic data, and think-aloud interviews all contribute to refining prompts and assessing how well the survey captures useful insights.

The \textbf{tool}, the AI-driven survey system, must be tested to confirm fair question presentation, appropriate adaptation, and avoidance of unintended response patterns. The iterative validation process includes analyzing both questions and responses to identify systematic biases or unintended influences from dynamic question generation. Refinements to the tool are informed by human responses, SQRA, expert evaluations, and think-aloud interviews, ensuring the system remains adaptive, trustworthy, and reliable.

\textbf{Rules} guiding validation include ethical and methodological considerations, differing for students and AI agents. For students, rules focus on informed consent, response privacy, and avoiding unintended influences that might bias answers. These rules align with best practices in survey methodology, e.g., \cite{LincolnGuba, walther}. Think-aloud interviews and expert reviews help validate question clarity and appropriateness, while member checking—inviting participants to review and respond to generated questions—enhances credibility and grounds the tool in real-world experience~\cite{LincolnGuba, Dorodchi}. Community standards define expectations for student engagement, including norms around survey fatigue, cognitive load, and response integrity. For AI agents in the SQRA, rules focus on controlling biases in synthetic responses, preventing hallucinations, and ensuring effective stress-testing of the survey instrument. These rules are established externally through researcher-defined constraints and internally through system prompts guiding AI behavior. Persona prompts shape AI responses, setting expectations for tone, depth, and interpretability. Additional constraints include model temperature, response randomness, acceptable variability, and benchmarking AI outputs against human responses to assess alignment with intended question objectives. By refining these prompts and constraints, AI-generated responses become a meaningful tool for evaluating and improving dynamically generated survey prompts. 

\textbf{Community} expectations shape validation methods, with standards set by both the broader academic community and the engineering education research community for survey design, response integrity, and AI usage. In SQRA, AI agents' ``community'' differs from human participants' and may influence their response behaviors and norms. AI agents lack societal and cognitive biases that may influence student responses, such as providing excessively detailed answers without the skepticism or hesitancy human participants might show. These prompts influence factors such as how many AI agents hesitate to engage with certain questions, how many exhibit reluctance or recalcitrance, and how reflections on teamwork are framed---whether they predominantly describe positive interactions or struggle with conflict resolution. Researchers must calibrate these prompts to create AI-generated responses that meaningfully contribute to survey validation, ensuring stress-testing without artificial agreement or over-generalized reflections.

The \textbf{division of labor} in validation involves multiple stakeholders: researchers design the validation framework and interpret results, expert reviewers assess question quality, students provide direct feedback on their survey experiences, and AI agents contribute to iterative refinement of both the survey instrument and the AI system through ``AI-in-the-loop" workflows. Researchers and expert auditors intervene to flag and revise problematic prompts, while AI agents conduct large-scale simulations to identify potential blind spots. This distribution of roles raises critical questions: How much automation should be used? When and where should human oversight intervene? Who should be responsible for reviewing, interpreting, and approving AI-generated content? By mapping these decisions through AT, we position validation not as a static step but as a socially and technologically mediated activity system, requiring ongoing negotiation of responsibility and trust to ensures dynamically generated questions remain effective, inclusive, and aligned with educational research goals.

\section{Methodology}

\begin{figure}[h]
\centering
\includegraphics[width=0.75\textwidth]{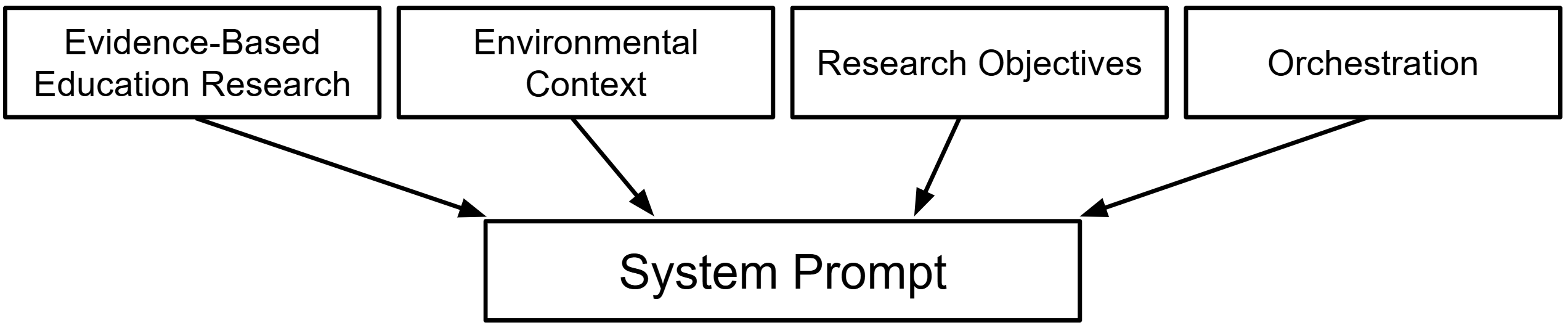}
\caption{System prompt structure for AI-generated questions and personas.}
\label{fig:systempromting}
\end{figure}

The methods section is divided into two components. First, we present a step-by-step guide for constructing an AI-driven, dynamic question generation system within the Qualtrics survey platform. This system uses OpenAI's GPT models to generate personalized, context-aware questions in real time based on participant responses. We also describe our prompt engineering strategies, which draw on evidence-based educational research, defined research objectives, environmental context, and orchestration techniques to support the generation of contextually relevant and pedagogically meaningful survey questions (see Figure~\ref{fig:systempromting}).

AT informed both persona and question-generation prompt design. Persona prompts were crafted to reflect realistic environmental and contextual factors---such as teamwork challenges and collaboration styles---aligned with the social norms participants might encounter. Question-generation prompts embedded situated participant perspectives, aiming to support reflective engagement grounded in constructs such as socially shared regulation of learning (SSRL)~\cite{hadwin2011self} and teamwork dynamics~\cite{adams_conceptual_2002}.

\vspace{-5mm}

\begin{figure}[h]
\centering
\includegraphics[width=0.5\textwidth]{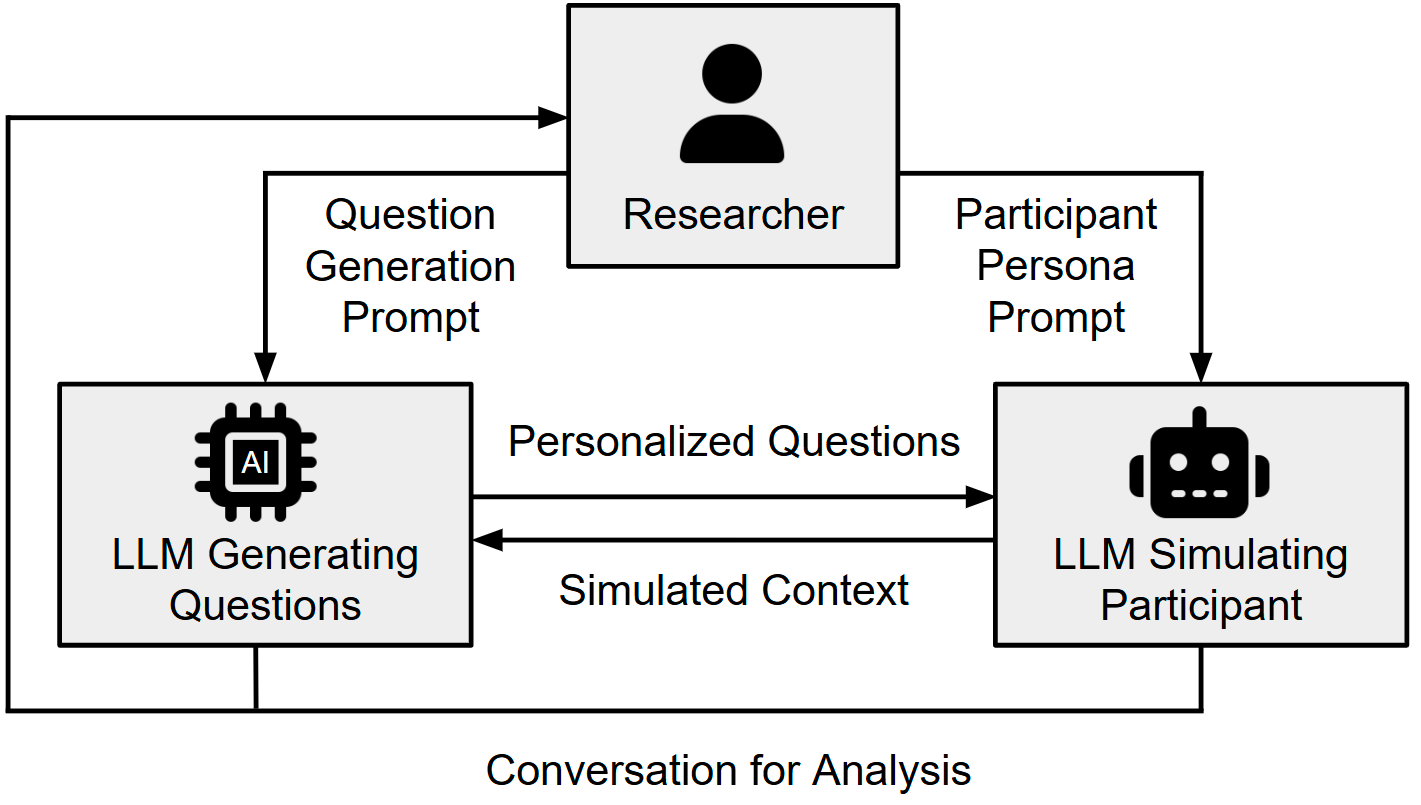}
\caption{Synthetic Question-Response Analysis (SQRA) framework.}
\label{fig:sqra}
\end{figure}

\vspace{-5mm}
Second, we introduce the SQRA framework (Figure~\ref{fig:sqra}), which uses simulated responses to begin validating AI-generated questions before deployment with human participants. We outline how SQRA is operationalized and describe our approach for evaluating how well synthetic testing replicates the results of human-tested surveys. This approach focuses not only on output quality but also on the potential for SQRA to serve as a scalable, lower-risk alternative for testing unbounded or untrusted AI systems prior to use  with human participants.

\vspace{-3mm}

 \vspace{-13mm}
\subsection{Survey Design and AI Integration} \label{sec:surveydesign}
\begin{figure}[h]
\centering
\includegraphics[width=0.6\textwidth]{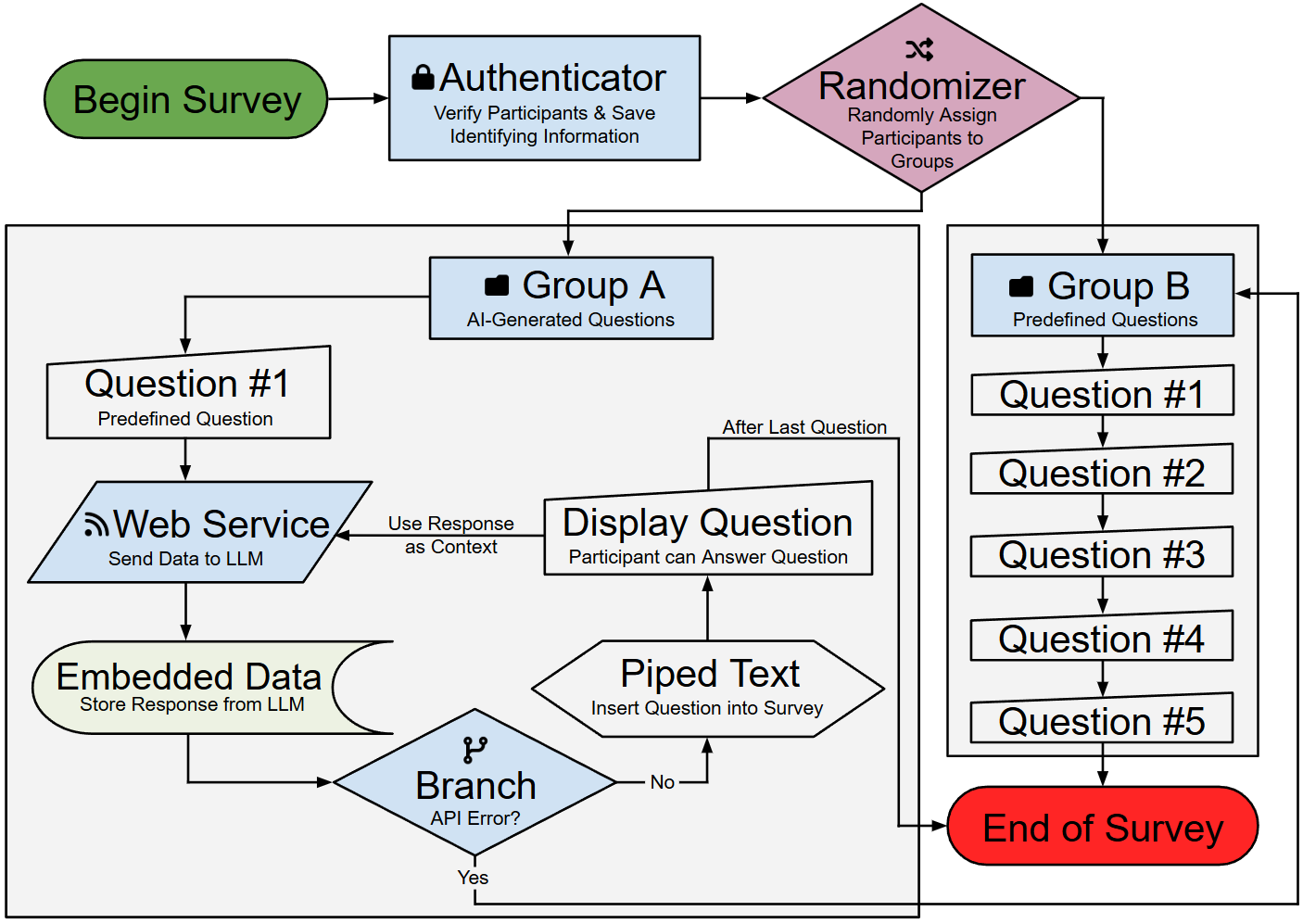}
\caption{Workflow diagram for integrating LLM into Qualtrics.}
\label{fig:figure1}
\end{figure}

Our survey was designed using Qualtrics, a widely utilized platform for creating, distributing, and analyzing surveys in both academic and commercial research. Figure \ref{fig:figure1}, is a workflow diagram that illustrates the integration of an LLM into the Qualtrics platform for dynamic question generation. It demonstrates the process by which participant responses are fed into the LLM through a Qualtrics Web Services, enabling real-time adaptive question generation. Users first authenticate their identity using their institution's login, facilitated by an Authenticator in Qualtrics. A Qualtrics Randomizer then assigns participants to one of two groups: Group A, which receives AI-generated questions, or Group B, the control group with predefined questions (see Appendix ~\ref{app:randomization} for more implementation details). For Group A, the first question is predefined by the research team. After the student responds, a Qualtrics Web Service sends the response to the LLM API, which generates a new personalized question or an error. The API response is stored as Embedded Data. A Qualtrics Branch checks the response for validity; if an error occurs, a predefined backup question is used. The Branch stores either the AI-generated or backup question as piped text, which is displayed to the participant. This process is repeated for a set number of questions before the survey concludes.

The first step in using the OpenAI API is obtaining an API key for authentication and security purposes. An API key is generated upon creating an OpenAI account. This key is essential for authenticating each API request made by Qualtrics. API keys also allow for access control, ensuring that only authorized users can interact with the OpenAI services.

Every API call to OpenAI includes the API key, which identifies the user and ensures that the request is valid. The API key enforces usage limits to prevent overuse. Through this, we are able to set limits on the number of requests per minute or per day, as well as the specific LLM model (e.g., GPT-3.5, GPT-4o) that can be accessed to manage costs. To ensure the security of sensitive information, it is important that the API key is not exposed to participants and no identifying data from participants is sent to the API, thus ensuring compliance with data privacy regulations and research compliance. For additional details on security and privacy, see Appendix ~\ref{app:securityandprivacy}.

In Qualtrics, you can hard-code JavaScript into your questions to make API calls to an LLM to create questions. While this method is functional, it poses a significant risk: when participants make API calls, your OpenAI key will be exposed. A workaround is to create a middle server that processes API requests while hiding your OpenAI key. However, this approach comes with challenges---the server must be highly secure to handle sensitive participant data and will incur additional operational costs. Further, it introduces another potential point of failure in the question-generation process. To avoid these complications, we recommend using the in-built Web Service feature in Qualtrics, which facilitates API calls while keeping your API key private and the participant data secure.

As participants answer questions, their responses are saved as embedded data (see Appendix~\ref{app:embeddeddata}), which informs subsequent questions generated by the AI. The system prompts used by the LLM to generate these questions are also stored as embedded data, ensuring that the interaction remains relevant and personalized. This setup allows the survey to adjust dynamically based on each participant's input. The embedded data is referenced throughout the survey using piped text (e.g., \verb|${e://Field/question2}|), enabling smooth and logical transitions between questions. In the event of an API error, the embedded data also stores backup questions to ensure the survey continues seamlessly; see Appendix~\ref{app:errorhandling} for further details on error handling. ChatGPT API errors are especially common during the release of new generative AI models---such as transitions from ChatGPT-3.5 to ChatGPT-4o---or the launch of new tools like SQRA. These upgrades or maintenance activities often result in temporary system downtime. By implementing this fail-safe mechanism, the survey continues to operate uninterrupted, allowing participants to complete the survey even during periods of system instability.

To enable the AI-driven question generation, Web Services in Qualtrics are used to integrate with OpenAI's API (see Appendix \ref{app:webservices}). Through this integration, Qualtrics sends requests to the OpenAI API whenever a new dynamic question needs to be generated. The responses provided by the participants serve as input data that the system forwards to the API. Based on the input, the API generates contextually relevant follow-up questions, which are then stored as embedded data and displayed in the following Qualtrics question. 

The AI-driven survey tool was deployed across multiple educational contexts to evaluate its adaptability and effectiveness in fostering engagement and reflection with STEM students. The first context was a one-credit introductory experimental physics laboratory course with approximately 400 students, primarily from physics and engineering disciplines, emphasizing teamwork and hands-on experimentation. The second context was a one-credit upper-level biomedical engineering experiential learning course with 54 enrolled and of those 38 students who agreed to participate in the research. More demographic data for these courses can be found in Appendix \ref{app:demographics}. In both courses, the survey tool was integrated into assignments requiring students to reflect on teamwork-related activities, while the reflection assignment was required for the course, participation in the research study was voluntary, and informed consent was obtained for research purposes. This study was approved by the IRB at
Cornell University (IRB0148748). All student responses were gathered in the Fall 2024 semester.

\subsection{The Synthetic Question-Response Analysis (SQRA) Framework}

The development of AI-driven survey tools introduces new challenges for assessing question quality, appropriateness, and alignment with research objectives. Traditional evaluation methods may not easily apply to adaptive, generative systems, particularly when the outputs vary in tone, relevance, or coherence. Unlike human-authored surveys, AI-generated questions can be unpredictable or reflect unintended biases, which may compromise participant trust---especially in sensitive or diverse settings~\cite{Kusters}. Moreover, the lack of transparency around how and why questions are generated complicates their use in applied research contexts.

\begin{figure}[h]
\centering
\includegraphics[width=0.6\textwidth]{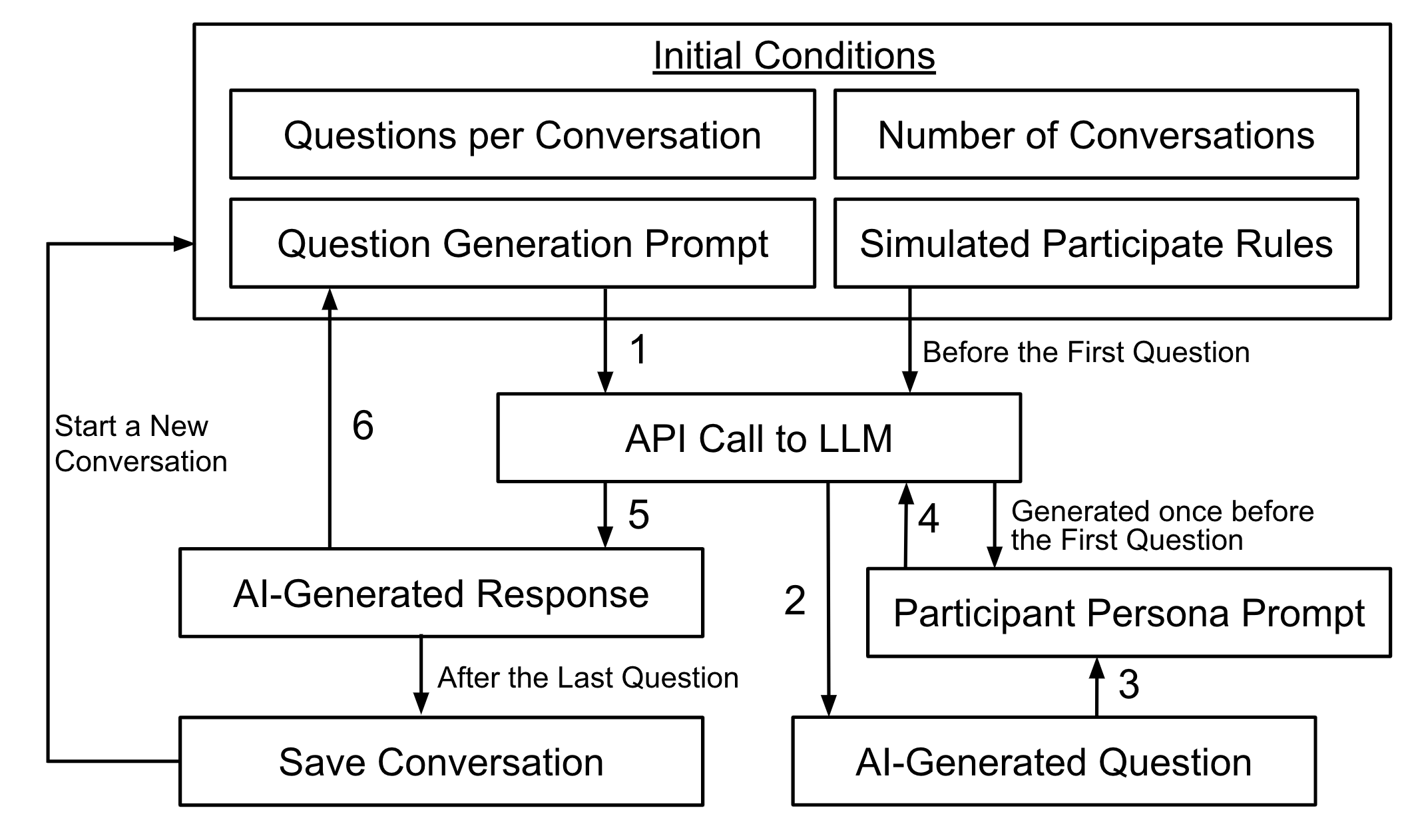}
\caption{Process of generating synthetic data with an LLM.}
\label{fig:figure8}
\end{figure}

To address these challenges, we introduce the SQRA framework, which allows researchers to test and refine AI-generated questions before using them with human participants. Figure~\ref{fig:figure8} illustrates the SQRA process, which consists of an iterative, multi-step interaction between synthetic participant responses and AI-generated questions, with Steps 1-6 repeated iteratively until the conversation concludes.

The process begins by defining key parameters, including: (i) the number of questions generated per conversation (set to 4 in this study); (ii) the total number of simulated conversations (1,000 in our case); (iii) the system prompt for question generation (see Section~\ref{sec:promptengineering-qualtrics}); and (iv) the prompt for generating a synthetic participant persona (see Section~\ref{sec:promptengineering-persona}). The participant persona prompt is used to simulate a consistent student perspective throughout the conversation.

Each synthetic conversation begins with a predefined initial question authored by the research team, shown in Step 1 of Figure~\ref{fig:figure8} (though this could also be AI-generated). This initial question and the persona prompt are submitted to the LLM, which returns a simulated student response. That response, along with the persona prompt, is subsequently fed back into the LLM API to dynamically create the next question. This process---feeding the generated question and participant persona prompt back into the LLM API---continues until the set question limit is reached for each conversation.

The SQRA framework enables the generation of large quantities of dynamic question-response pairs that can be reviewed and analyzed prior to human deployment. However, a critical question remains: how well does the SQRA framework replicate the types of questions generated in human-to-AI interactions (\textbf{RQ2})? To evaluate this, we compare the characteristics of questions generated under synthetic and human conditions through lexical and grammatical analysis (see Section~\ref{sec:validating}).

\subsection{Prompt Engineering}
The process of prompt engineering was crucial in guiding the AI to generate contextually relevant and educationally aligned questions. To achieve this, prompts were carefully designed to reflect findings from educational research. The prompts were also contextualized to the learning environment of the course and provided with specifications for the desired tone and style of the AI agent's responses (i.e., ``orchestration''), see Figure~\ref{fig:systempromting}.

\subsubsection{Question Generation Prompt} \label{sec:promptengineering-qualtrics}

For question generation, our prompt was specifically designed to support the goals of fostering reflection and assessing the use of SSRL ~\cite{hadwin2011self}. An example of our question generation prompt is provided in Box \ref{text:questiongen}. The prompt integrates contextual information about the courses, the objectives of the prompts, relevant research on SSRL, e.g., ~\cite{hadwin2011self} and effective team dynamics, e.g., ~\cite{adams_conceptual_2002}, as well as orchestration details including: (i) asking only one open-ended question at a time, (ii) acting as a critical and professional coach, and (iii) maintaining a supportive, encouraging, and neutral tone throughout the conversation.

AT provides a structured framework that helps guide the development of the question-generation prompt by ensuring that the prompt takes into account all the components of the activity system. The subject is the human participant with diverse experiences, perceptions, and engagement levels. The tool is the AI-driven survey system, which is designed to facilitate engagement. The object is the participant responses that are collected by the researcher that in a future work, we will explore further. The rules are the ethical, cultural, and contextual norms that shape human engagement. The community represents the social and educational environments of the participants, influencing the emphasis on teamwork, collaboration, and effective communication in the prompt design. The division of labor involves the human participant's role in actively contributing reflective responses while the AI system facilitates the process by generating potentially contextually relevant questions. By grounding the prompt design in these AT components, researchers ensure the prompt is comprehensive, contextually relevant, and effective in meeting the study's goals.

The use of team performance coaching as an example illustrates how AI-integrated surveys can identify specific dynamics, such as collaboration challenges or individual accountability, which are critical constructs in educational research. While this example focuses on team coaching, it serves as a proxy to explore the adaptive capabilities of AI-integrated surveys for generating contextually relevant questions. AI-driven surveys present a novel approach to data collection by tailoring questions to participant responses. For instance, researchers could use AI-driven surveys to collect data from students about their experiences in team-based projects, dynamically adapting to nuanced interactions. This adaptability highlights the versatility of the framework, underlining its broader applicability across diverse research contexts and its potential to advance survey-based research methodologies by capturing richer, more nuanced data.
\vspace{20mm} 
\begin{boxwithhead}
{BOX 3.3.1\quad   Example Question Generation Prompt \label{text:questiongen}} 
{\noindent You are a critical and professional coach for students taking introductory physics classes at a University. Students are working in small teams of 4 to 5 people to complete lab activities and related assignments. Your goal is to engage students in reflective conversations about their teamwork experiences, fostering critical thinking, metacognitive skills, and promoting deeper cognitive processing. 
\vspace{2mm} 

Read the student's response and reflect on each step of the conversation to decide what to ask next. Ask only 1 open ended question at a time.
\vspace{2mm} 
Instructions:

{\bf Questioning:} Only ask a question or seek clarification. Do not provide new information. Make sure all questions can help students to grow and reflect on their own experiences. Make sure all questions are related to teamwork and student's experiences. If a student's response is too short and simple, ask for deeper reflective questions.

{\bf Conciseness:} Keep your responses friendly, short and focused. Avoid jargon or terms unfamiliar to students to ensure clear communication.

{\bf Guidance:} Make sure all questions are related to teamwork and building a smooth and healthy teamwork environment. Frame all your questions around the following collaborative learning principles but do not use the terms directly: 

{\bf Principle 1:} Positive interdependence: an individual's success is connected to the group's success; 

{\bf Principle 2:} Individual accountability: each member is responsible for their contribution to the team effort; 

{\bf Principle 3:} Face-to-face promotive interaction: a group working together directly and supporting one another's efforts to solve problems; 

{\bf Principle 4:} Social skills: includes skills like communication, conflict resolution, leadership, decision-making, and trust-building; 

{\bf Principle 5:} Group processing: reflecting on a group session to describe what member actions were helpful or unhelpful and making decisions about what behaviors to continue or change.

{\bf Deepening Reflection:} Encourage students to reflect on their engagement with socially shared regulation of learning (SSRL). SSRL involves the group collectively negotiating and aligning their perceptions of the collaborative learning process, and taking control of the task through shared, iterative fine-tuning of cognitive, behavioral, motivational, and emotional conditions. Guide them to discuss how they collectively negotiate, align their perceptions, and take control of their learning process through iterative adjustments

{\bf Tone:} Maintain a supportive, encouraging, and neutral tone throughout the conversation.}
\end{boxwithhead}

\subsubsection{Participant Persona Prompts}  \label{sec:promptengineering-persona}

The ``Simulated Participate Rule,'' shown in Box \ref{text:persona}a, serves as a set of guiding rules that shape the behavior and characteristics of our \emph{in silico} student. Drawing on educational research, we crafted this prompt to reflect common challenges students face within team settings---such as unequal workload distribution, scheduling conflicts, and differences in work styles~\cite{blumenfeld1996learning, van2006social, volet1998culturally, education2014211A3, SALOMON198989}---as well as typical student reflection styles~\cite{franklin2018improving, may2002college, werth2023investigating, treibergs2022introductory, csavina2016assessing}. This prompt is intended to create simulated students who closely emulate our target population, allowing us to evaluate the types of questions generated by our survey tool within the SQRA framework.
\vspace{20mm} 
\begin{boxwithhead}
{BOX 3.3.2a\quad  Participant Persona Prompt \label{text:persona}}
{\noindent Generate a persona for a student in a scenario that reflects common teamwork successes or challenges encountered in STEM higher education. This student will answer survey questions tailored to their experiences. Their responses will be analyzed by researchers, and educators. Your persona should be detailed and provide context around the student's background, motivations, and experiences.

Your response should follow this structure:

{\bf Background:} This student is a Biomedical Engineering major. Describe their personal characteristics (e.g., age, race, gender, socioeconomic status, expertise, level of academic confidence, underrepresented minority in STEM status, first-generation college student, returning adult learner, international or domestic student, etc.). 

{\bf Team Dynamics:} Describe their role in the team and the teamwork dynamics. Are they a leader, contributor, or an observer? Detail whether the student is experiencing challenges (e.g., teammate tardiness, unequal workload distribution, scheduling conflicts, work style clashes, lack of communication, etc.) or working in a highly functional team. Specify if there are additional barriers, such as imposter syndrome, feelings of isolation, stereotype threat, or inequity in group role assignments.

{\bf Motivations and Engagement:} Outline the student's motivations for studying in their field (e.g., interest in research, desire to innovate, societal impact, specific career goals) and their perspective on teamwork. Include whether they see teamwork as an essential skill, a frustrating barrier, or a mix of both. How much is the student enjoying the class? Determine how engaged this student will be (e.g., mostly one-word responses vs short sentences vs thoughtful insights) with the survey questions they receive. How comfortable is this student with AI and other technology? What are their personal views on AI and do they enjoy using it? Do they see taking surveys as a waste of time?

{\bf Reflection Style:} Describe the student's approach to reflection. Are they inclined to reflect deeply on their experience? Will they respond with thoughtful insights, short phrases, or single words? Indicate any tendencies toward genuinely monitoring their thought patterns and learning processes, or focusing more on meeting assignment requirements without deep engagement.

Be concise.}
\end{boxwithhead}

AT informs the development of the ``Participant Persona Prompts'' and the ``Generating Student Responses Prompt'' by guiding researchers in aligning the prompts with the components of the activity system. The subject is the simulated AI persona, designed to emulate the characteristics and behaviors of human student participants. The tool is the AI-generated prompts that guide the creation of these personas and their reflective responses, ensuring they authentically represent real-world scenarios. The object is the generation of participant responses that reflect teamwork successes and challenges, enabling researchers to evaluate the appropriateness, adaptability, and potential biases of AI-generated questions. The rules include the structured format and ethical considerations for creating realistic personas and responses while grounding in educational research. The community represents the broader educational and social contexts influencing the personas, ensuring that simulated responses authentically reflect the dynamics of real student teamwork and collaboration informed by past education research. The division of labor reflects the AI's role in generating personas and responses and the researchers' role in validating these outputs to align with the study's goals. By addressing these components, AT ensures that the persona and response prompts support the dynamic survey's ability to generate meaningful and realistic data for analysis.

Our objective is that through these simulated responses, researchers can assess the appropriateness, adaptability, and potential biases in the AI-generated questions, ensuring that they align with the educational objectives of the study. This preliminary testing phase is crucial for validating that the questions foster meaningful reflection and engagement and that they are suitable for a wide range of student experiences.

The ``Simulated Participate Rules'' prompt is referenced during the SQRA framework prior to the generation of the first question to create the ``Participate Persona Prompt`` (shown in Figure~\ref{fig:figure8}). We use the prompt shown in Box~\ref{text:generating}b to create the persona that is used throughout a single simulated conversation. This prompt focuses on ``orchestration'' aspect of the student response to emulate the prose of a typical student reflective response. 

\begin{boxwithhead}
{BOX 3.3.2b\quad Generating Student Responses Prompt \label{text:generating}}
{\noindent You are the student. You will be asked questions about your experiences.
Keep responses in paragraph form and avoid using numbered lists or bullet points.}
\end{boxwithhead}

\subsection{Comparing SQRA Framework to \emph{in situ} Application of AI-Generated Questions} \label{sec:validating}

\begin{figure}[h]
\centering
\includegraphics[width=\textwidth]{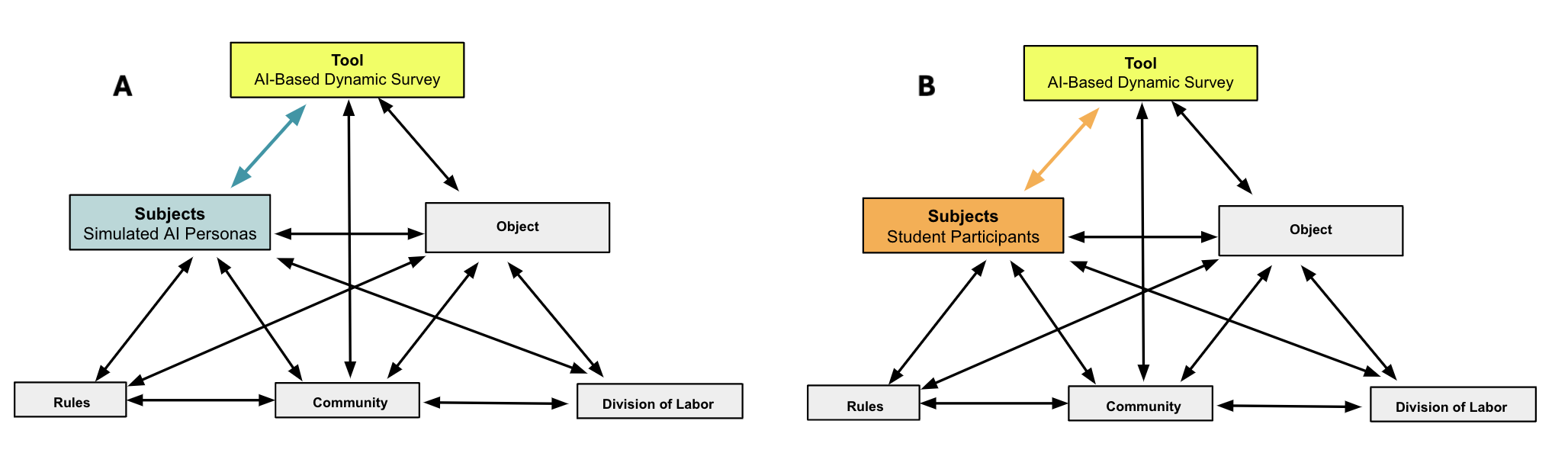}
\caption{Activity Theory triangles comparing simulated AI and human participants as the subjects.}
\label{fig:at2}
\end{figure}

Our results focus on comparing the SQRA framework to the \emph{in situ} application of an AI-generated survey tool. Grounded in AT, Figure~\ref{fig:at2} illustrates how shifting the subject---from a simulated AI persona (\textbf{A}) to a human participant (\textbf{B})---alters the components and interactions within the AI-driven survey activity system. This comparison explores how altering the subject affects the interactions and outcomes within the activity system. 

To compare these dynamics and address \textbf{RQ2}---how well the SQRA framework replicates the types of questions generated in human-to-AI interactions---we employ several analytical techniques: (i) sentiment analysis, (ii) cosine similarity, and (iii) question structure analysis. These metrics allow us to compare the types of questions generated in AI-to-AI versus AI-to-human interactions, as well as to identify and examine any outlier questions. 

Violin plots are used to visualize the distribution of AI-generated survey responses, as they combine box plots with kernel density estimation to show both summary statistics and response variability. Unlike traditional box plots, violin plots reveal the shape, spread, and density of responses, making them ideal for comparing AI-to-AI versus AI-to-Human interactions. The width of each violin represents response density, allowing us to detect patterns such as skewness, clustering, or multimodal distributions. This approach helps us understand response behavior, identify biases, and evaluate how AI-generated responses compare to human-administered surveys.

By framing this analysis within AT, we systematically explore how changes in the subject influence the broader activity system. While this paper primarily focuses on the subject, tool, and object, future research will delve into the roles of rules, community, and division of labor. This foundational work lays the groundwork for broader studies into the complexities of \emph{in silico} environments, confounding variables, and their implications for dynamic and scalable research tools.

\subsubsection{Sentiment Analysis}
Sentiment analysis, also known as opinion mining, is one of the most widely applied techniques in natural language processing~\cite{mite2018sentiment, wankhade2022survey, cambria2017practical}. This technique is used to assess the attitudes and tone embedded within textual responses~\cite{wankhade2022survey}. In this study, we apply sentiment analysis to evaluate the tone of AI-generated questions, categorizing them as positive, negative, or neutral. Additionally, we calculate a compound sentiment score by summing the valence scores (positive, negative, and neutral) and normalizing the score to a scale from $-1$ (most negative) to $+1$ (most positive).

We use the $SentimentIntensityAnalyzer$ from the Natural Language Toolkit (NLTK) library, along with the VADER lexicon. NLTK is a widely used, open-source library for natural language processing (NLP) in Python~\cite{Hanna2023AssessingBias}. The VADER lexicon is an extensive, transparent, and reliable sentiment analysis tool, validated through numerous human assessments~\cite{Bonta_Kumaresh_Janardhan_2019}. Its user-friendly interface and adaptability to a wide range of tasks made it suitable for our text analysis needs. By assessing the emotional tone in the text, we gained insights into the attitudes, opinions, and emotions expressed by the AI-generated open-ended survey questions.

To determine whether there was a significant difference in tone between AI-to-AI and AI-to-Human interactions, we conducted a Mann-Whitney U test~\cite{mcknight2010mann}. Additionally, we calculated Cohen's $d$ to measure the effect size, providing insight into the magnitude of differences between groups, where small $(d = 0.2)$, medium $(d = 0.5)$, and large $(d \geq 0.8)$~\cite{sullivan2012using, becker2000effect}.

\subsubsection{Cosine Similarity Analysis}
Cosine similarity, a metric for evaluating text similarity, ranges from 0 to 1, where 0 indicates no similarity between text samples and 1 represents perfect similarity~\cite{park2020methodology}. We applied cosine similarity to assess the similarity between each dynamic question and the preceding participant response within both AI-to-AI and AI-to-human interactions.

To analyze the conversational data, we used the $spaCy$ library for language processing and similarity measurement. Specifically, we loaded the $ \textit{en\_core\_web\_lg} $ in $spaCy$ model to perform semantic similarity calculations between text entries~\cite{Hanna2023AssessingBias}. 

A Mann-Whitney U test~\cite{mcknight2010mann} was conducted to determine if there were significant differences in cosine similarity scores between AI-to-AI and AI-to-Human interactions. We also calculated Cohen's $d$ to measure the effect size of these differences~\cite{sullivan2012using, becker2000effect}.

\subsubsection{Question Structure Analysis}
To evaluate the alignment between the AI instructions, provided through system prompts, and the appropriateness of generated questions, their lexical and grammatical structures were analyzed. Our AI tool was instructed to only ask one question at a time and not reply with comments, the frequency of conjunctions (e.g., but, and) and punctuation usage (e.g., question marks, commas, periods) were quantified as a proxy for doubled-barreled questions and alignment with AI prompting. This was then compared between AI-to-AI and AI-to-Human interactions to further understand how employing the SQRA framework changes question structure. Paired t-tests~\cite{hsu2014paired} were used to determine if frequency of conjunction and punctuation usage significantly differed between AI-to-AI and AI-to-Human interactions.

\subsection{Participant Populations and Course Contexts}
The Qualtrics survey tool was piloted in two courses: a one-credit introductory experimental physics laboratory course with approximately 400 physics and engineering students, as well as a third-year, one-credit seminar with 55 biomedical engineering students. Both courses feature significant amounts of teamwork and reflection. Instructors integrated the survey into course assignments, requiring students to reflect on teamwork-related activities for credit. Students were asked to consent to the use of their course artifacts for research purposes, with assurances that participation was voluntary and would not affect their course performance or relationship with the instructor following institutional review broad approval from partnering universities. No demographic data was collected for this study.

To evaluate the survey tool, we conducted an analysis comprising 1,000 AI-to-AI interactions generated through the Synthetic Question-Response Analysis (SQRA) framework and 318 AI-to-human interactions collected during the courses. Each interaction began with a predefined question, followed by four AI-generated questions. In total, we analyzed 4,000 AI-generated questions paired with AI-generated responses and 1,272 AI-generated questions paired with human-generated responses.

\subsection{Positionality of the Researchers}
The first author (TKM) is a graduate engineering student specializing in engineering education who developed the AI research tool, analyzed data, and contributed substantially to the writing of this paper under the mentorship of the last author. The last author (AW) is an Assistant Professor of Biomedical Engineering and Engineering Education Researcher. She holds a Ph.D. in Electrical and Computer Engineering and has extensive experience in engineering and physics education, as well as in applying AI methods in scientific research, with more recent work extending into educational contexts. The research team also includes a biomedical engineering undergraduate student (KR) who continued the development of the research tool, dissemination to participants, and data analysis as well as an engineering education postdoctoral researcher (CJM) who assisted with the AI tool development and writing of this paper.

We recognize that AI, particularly GenAI, is an evolving field with transformative potential. As researchers, we are excited about the opportunities it offers to scale research methods and enhance educational tools. However, we remain mindful of the significant challenges and ethical concerns this evolution brings, including issues of equity, trust, accuracy, and the potential for job displacement. We believe that responsible advancement in AI requires increased regulatory measures and a commitment to transparency in the development and application of these technologies. We advocate for increased oversight and open disclosure of the methodologies employed in AI to ensure its benefits are accessible, fair, and aligned with our values.


\section{Results}

The successful deployment of this AI-driven survey tool through Qualtrics in two separate courses further validates its utility and stability in practical, educational settings. Integration with Qualtrics was smooth, with no technical issues reported, indicating that the tool is compatible with the student's existing technology. 

\subsection{Sentiment Analysis Findings}

\begin{figure}[h]
\centering\includegraphics[width=0.8\textwidth]{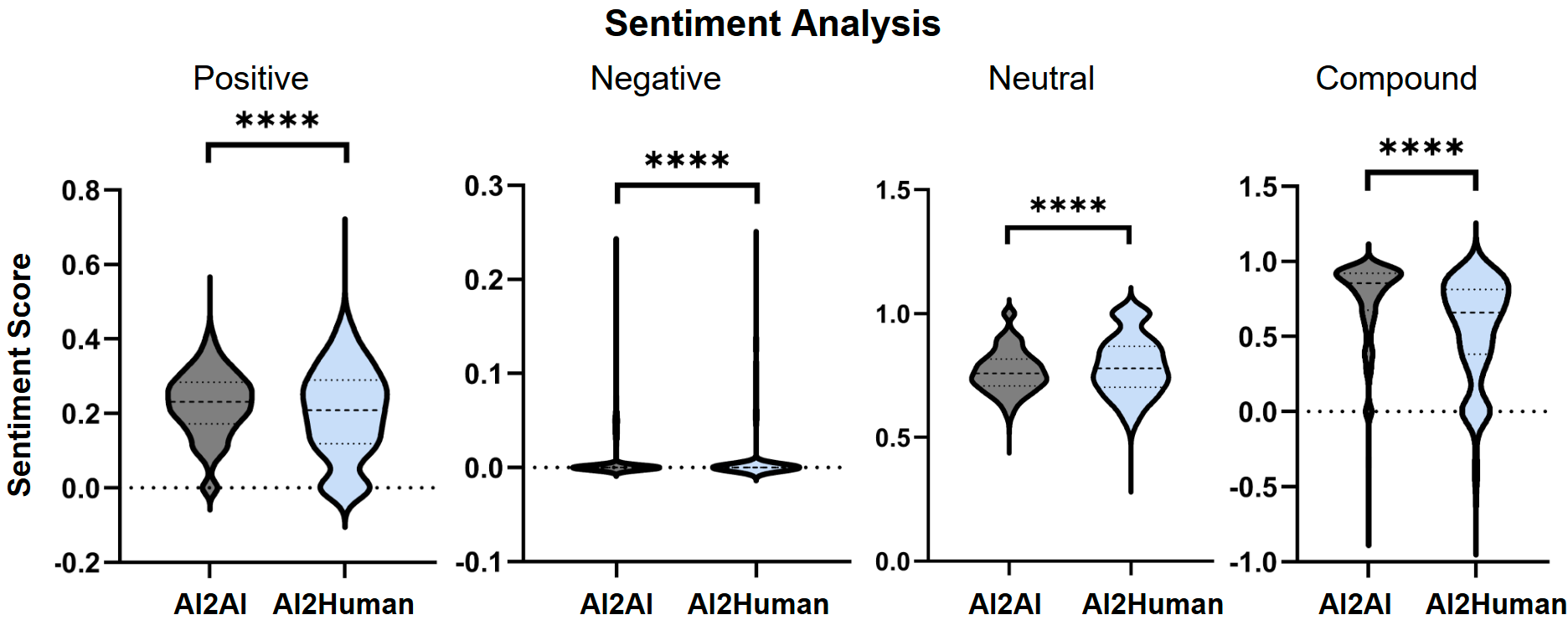}
    \caption{Sentiment scores for AI-to-AI and AI-to-Human interactions.} \label{fig:sentimentAnalysis2}
\end{figure}

The distribution of sentiment scores across AI-to-AI and AI-to-Human interactions was analyzed to determine the nature of responses. As shown in Figure \ref{fig:sentimentAnalysis2} and Appendix Figure \ref{fig:sentimentAnalysisPosNegNeut}, the majority of responses in both interaction types were neutral, with notable positive outliers and minimal negative sentiments. The violin plots indicate statistically significant differences between AI-to-AI and AI-to-Human sentiment scores. Effect size analysis, shown in Appendix \ref{app:effectsize}, demonstrated that the negative sentiment score had an almost zero effect, while the positive and neutral scores showed small effect sizes. The compound sentiment score exhibited a moderate effect size, suggesting that AI-to-Human interactions tend to have more neutral tones compared to AI-to-AI responses. In the figure, **** notation signifies a $p$-value $<0.0001$ for a Mann-Whitney U test.

Using sentiment analysis, we can identify and explore sentiment outliers within our dataset to understand the range and nuances in the types of questions being generated. These outliers reveal key insights into the nature of interactions, as well as the tone differences between AI-generated and human-targeted questions.

For instance, the most negative sentiment in the AI-to-AI interaction, \emph{``How did openly discussing the workload distribution with your teammates impact your feelings of isolation and frustration within the team?''} (Compound Score: -0.700), suggests that AI-to-AI questions can sometimes address emotionally complex topics. Similarly, in the AI-to-Human interactions, a negative sentiment outlier, \emph{``In reflecting on the team's process, how did your group handle disagreements or conflicting ideas when troubleshooting the code?''} (Compound Score: -0.586), demonstrates the model's ability to prompt critical reflection on challenging team dynamics. Although both questions convey a slightly negative tone, they invite constructive discussions on problem-solving and conflict resolution. Notably, the AI-to-Human question also contextualizes the teamwork challenge within a specific activity, ``coding,'' relevant to students' experiences.

On the other hand, the most positive AI-to-AI interaction, \emph{``That sounds like a thoughtful approach to improving teamwork dynamics. How do you think clearly defining roles and responsibilities can help us build stronger positive interdependence within our team?''} (Compound Score: 0.962), demonstrates that the model occasionally provides evaluative feedback, even when not intended. This response contrasts with our system prompt's design, which specified single-question prompts without judgments or statements. Furthermore, this question includes the term ``positive interdependence,'' an educational term that might be unfamiliar to students.

Finally, the most positive AI-to-Human interaction, \emph{``How do you think each team member's strengths and roles contribute to the overall success of the team?''} (Compound Score: 0.751), is similar to the two negative outliers but framed in a more positive way. In all these cases, the questions are double-barreled (i.e., they ask about multiple topics at once but allow only a single answer), which could lead to less interpretable data depending on the research objectives. 

\subsection{Cosine Similarity Analysis Findings}

\begin{figure}[h!]
\centering
\includegraphics[width=0.5\textwidth]{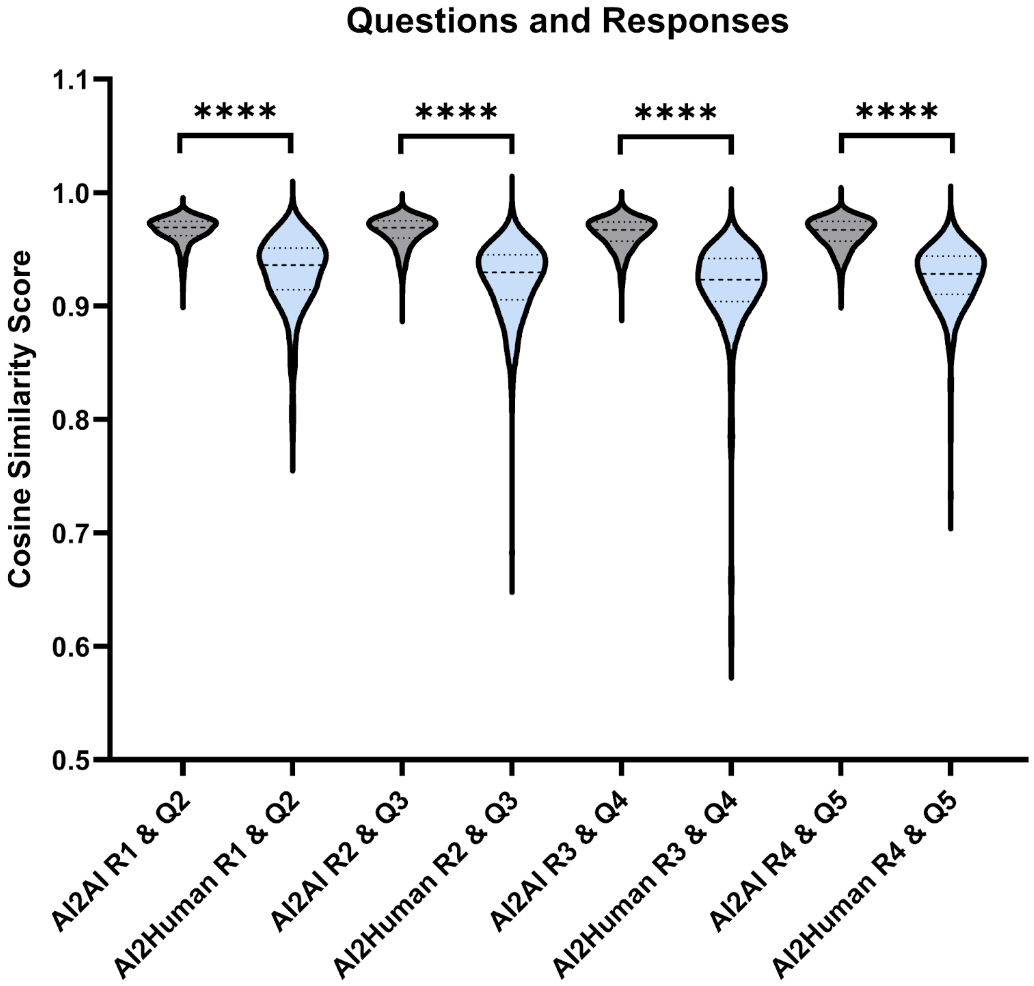}
\caption{Cosine Similarity Scores Comparing AI-to-AI and AI-to-Human Interactions. }
\label{fig:cosineSimilarity}
\end{figure}

Cosine similarity was used to understand how the AI-generated question was responding to the subject's previous input. We compare the results from the AI-to-AI and Human-to-AI for each response-question pair shown in Figure \ref{fig:cosineSimilarity}. The **** notation signifies a $p$-value $<0.0001$ for a Mann-Whitney U test. Overall, we see strong similarities ($\geq 0.5$) for both populations. However, there are still substantial, significant differences between the two groups with the Human-to-AI group showing lower similarity scores, more variance, and more outliers. The effect size analysis, detailed in Appendix \ref{app:effectsize}, provides numerical support for these findings. We examined the text of sentiment and similarity outliers to explore differences between AI-to-AI and AI-to-Human interactions, with full text examples provided in Appendix \ref{app:cosinesimtext}. 

In this analysis, we observed that all AI-to-AI interactions had high similarity scores (0.99 is the highest and 0.90 is the lowest). These as well as the AI-to-Human example with the highest sentiment score of 0.97, contain long, elaborate responses. For instance, the \emph{in silico} persona with the highest cosine similarity score provided the following reflection, along with the AI agent's follow-up question:

\begin{quote}
\textbf{Simulated Student:} Reflecting on the team dynamics compared to last week, I notice a mix of progress and ongoing challenges. In terms of collaboration, there was a slight improvement as some members made an effort to involve everyone in discussions, allowing for more balanced participation. I appreciated this shift, as I felt more included in the brainstorming process and was able to share some of my ideas about our software solution. However, there were still moments when a few individuals dominated the conversation, leaving little room for others to contribute. The scheduling conflicts we faced last week continued to be an issue this week, making it difficult for everyone to be on the same page. This ongoing struggle has created an undercurrent of frustration, as it often feels like we are not fully capitalizing on everyone's strengths, including my own. While I sense some movement toward better teamwork, I still find myself navigating feelings of isolation and the burden of unequal workload distribution, which makes me even more determined to advocate for a more inclusive team dynamic moving forward.

\vspace{1em}

\textbf{AI-generated Question Response:} It's great to hear that there was some progress in including everyone in the discussions this week. It seems like you are feeling a mix of improvements and ongoing challenges in your team dynamics. Can you tell me more about how you plan to advocate for a more inclusive team dynamic moving forward, especially in addressing the concerns of domination in conversations and unequal workload distribution?

\end{quote}

While the follow-up question is specific, contextualized, and invites deeper reflection, it includes two sentences summarizing and evaluating the response---an approach that deviates from our system prompt's guidelines to avoid evaluative feedback.

A similar pattern emerged in the AI-to-Human interaction with the highest cosine similarity score. The student's response was detailed, and the AI-generated follow-up built directly on that reflection. Though it includes two follow-up sentences, both are specific and aligned with the student's experiences. The full exchange—student reflection followed by the AI's question—is shown below:

\begin{quote}
\textbf{Student:} To be fair, I know there is much to improve in my strategies to get them to refocus on the lab. I either worked through the report myself, or if I had questions I interrupted their conversation to ask for help. Sometimes I did have to ask multiple times before I got them to focus on my question. To their credit, they did offer great help once I did get them to refocus. Towards the last 30 minutes of our lab I started to be more assertive and push them to work since I got pressured by the time. I think by being more assertive, that helped them focus more on the lab. Ultimately however, they did not have the motivation to progress through the lab and I had to push them to complete it and be more ambitious with the experiments we were conducting.

\vspace{1em}

\textbf{AI-generated Question Response:} How did you react when you had to push your teammates to work on the lab while you were typing up the report? How did that make you feel about your role within the team?
\end{quote}

These examples contrast sharply with the Human-to-AI outlier with the lowest cosine similarity score, where the student simply wrote, \emph{``True,''} to which the AI survey instrument responded, \emph{``That sounds like a positive change! How did being more responsive to each other's ideas impact the overall effectiveness of your team's work?''} These brief responses accounted for most of the lower similarity scores in our analysis.

\subsection{Question Structure Analysis Findings}
\begin{figure}[h]
\centering
\includegraphics[width=1\textwidth]{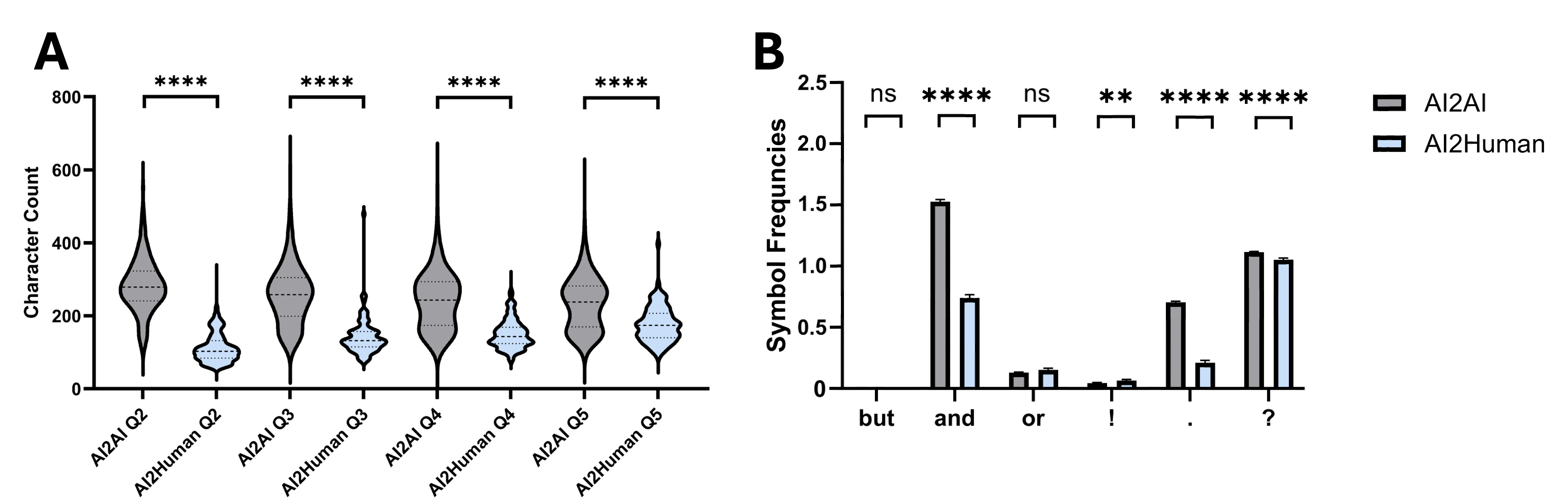}
\caption{Structural comparison of AI-generated questions.}
\label{fig:CCnSF}
\end{figure}

We investigated various structural elements of the AI-generated survey questions, including character count and symbol frequency. Figure~\ref{fig:CCnSF}A highlights substantial differences in question length between AI-to-AI and AI-to-Human interactions. Specifically, the survey instrument generated longer questions in AI-to-AI settings than in AI-to-Human settings, with statistically significant $p$-values and large effect sizes (see Appendix~\ref{app:effectsize}). These findings focus exclusively on the questions generated by the AI-based survey tool—not on the responses written by student participants or simulated AI personas—and suggest that the instrument tends to produce lengthier questions when interacting with AI agents, while questions directed at human participants are typically more concise. Appendix Table~\ref{tab:characterCountQuestions} provides examples of long and short AI-generated questions across both interaction types, illustrating this contrast in length.

Figure \ref{fig:CCnSF}B shows the frequency of specific symbols and words used per question. AI-to-AI responses showed a higher frequency of ``and'' and periods, likely reflecting their longer structure. In contrast, other symbols and words, such as ``but,'' ``or,'' and ``!'' show minimal differences between AI-to-AI and AI-to-Human interactions.  Additionally, question marks appeared on average once per question in both interaction types, indicating consistent alignment with the system prompt instructions. The ****, **, and $ns$ notations indicate a $p$-value $< 0.0001$, a $p$-value $< 0.01$, and no statistical significance, respectively, based on a Mann–Whitney U test for \textbf{A} and a paired $t$-test for \textbf{B}. Error bars represent the standard error of the mean.

\vspace{-5mm}

\subsection{System Prompt Refinement}
We analyzed AI-generated survey questions from both AI-to-AI and AI-to-Human interactions to identify common issues in question quality. Across both conditions, four recurring problems emerged. First, the AI often used repetitive language and redundant phrasing, which reduced engagement and limited the depth of participant responses. Second, many questions were overly long and wordy, potentially overwhelming respondents. Third, AI-generated questions frequently included evaluative or affirming language---such as \emph{``That sounds like a very thoughtful approach...''}---which introduced unintended bias. Finally, the tool regularly produced double-barreled questions (e.g., \emph{``Can you share how you motivated disengaged teammates? How did you handle that challenge?''}), which may confuse participants and complicate analysis.

To improve question quality, we applied the SQRA framework through iterative AI-to-AI simulations. Following Walther et al.'s quality framework for qualitative research~\cite{walther, walther2013}, we evaluated each round of questions using synthetic student responses and aligned revisions with three core validation principles: theoretical, communicative, and procedural. Theoretical validation centers on generating trustworthy knowledge and fostering understanding. Here, precise wording and linguistic diversity increase clarity. Communicative validation emphasizes shared meaning-making across audiences, requiring accessible phrasing that avoids unnecessary length. Procedural validation calls for minimizing random variation while promoting open-ended responses through a flexible interview structure.

Based on these principles, we refined the system prompt. Updated prompting guidelines included: \emph{``Ask only one question at a time,'' ``Avoid double-barreled questions,''} and \emph{``Do not begin with affirmations or evaluative comments (e.g., `That sounds like…').''} We also introduced flow guidance such as: \emph{``If the response is already detailed, shift to another learning principle to avoid redundancy.''} Tone instructions emphasized clarity, conciseness, and student-centered phrasing: \emph{``Use precise, meaningful language and promote thoughtful exploration of teamwork dynamics.''}

\begin{figure}[h]
    \centering
    \includegraphics[width=1\textwidth]{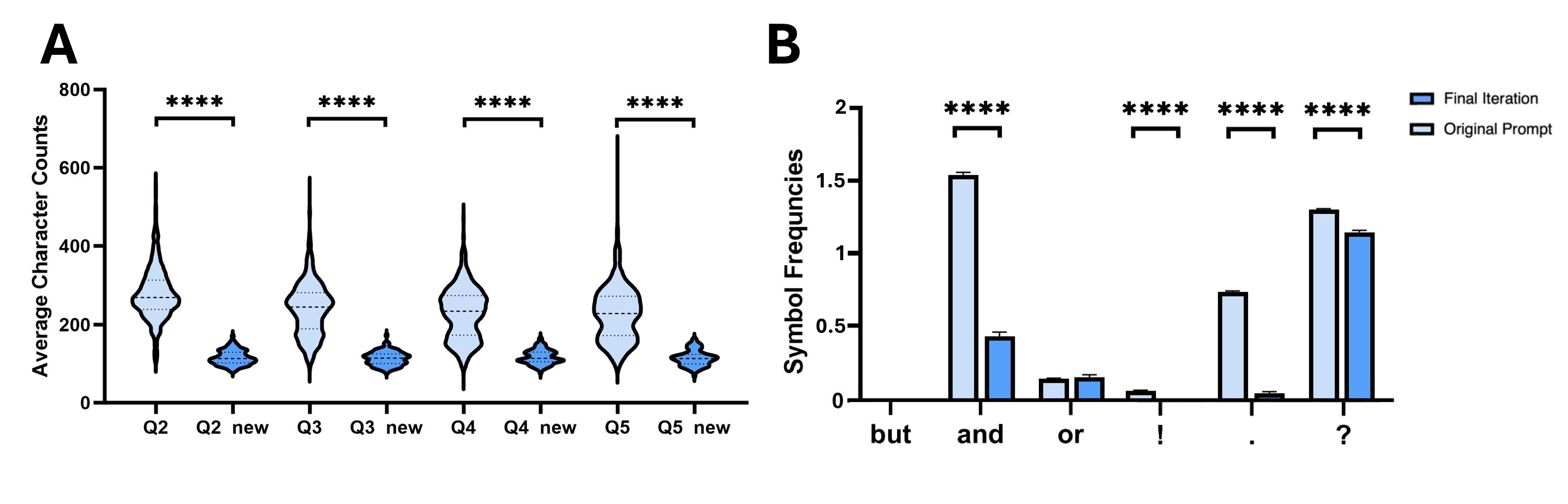}
    \caption{Impact of prompt refinement on question length and structure.}
    \label{fig:SQRA_refinement}
\end{figure}

Refinements were implemented iteratively across five cycles, modifying one prompt instruction per cycle and simulating 100 AI-to-AI student conversations each round using the SQRA framework. Compared to the original prompt, the final iteration produced shorter, more focused questions (Figure~\ref{fig:SQRA_refinement}A) and fewer sentences per question, as indicated by reductions in periods, conjunctions, and question marks (Figure~\ref{fig:SQRA_refinement}B). Revised questions avoided implicit evaluations and instead emphasized actionable, specific strategies. 

As part of the system prompt refinement process, we updated the system-prompting AI from ChatGPT-3.5 to GPT-4o. GPT-4o’s enhanced adaptability, tone modulation, and contextual relevance led us to update both the student AI and the system prompt accordingly. We observed compounding improvements resulting from both the prompt refinements and the model upgrade. This also highlights SQRA's utility in evaluating the impact of frequent model updates as GenAI capabilities rapidly grow and improve.

Figure~\ref{fig:SQRA_refinement} compares questions generated in the final refined iteration to those from the original prompt. Panel \textbf{A} shows changes in character count; panel \textbf{B} shows symbol frequency. **** denotes a $p$-value $< 0.0001$, based on a Mann–Whitney U test for \textbf{A} and a paired $t$-test for \textbf{B}. Error bars represent the standard error of the mean.

\section{Discussion} \label{sec:discussion}

AI-driven surveys fill a unique role in education research by combining the scalability of traditional surveys with the personalized engagement typically seen in interviews. Traditional surveys, while efficient for large-scale data collection, often lack the flexibility to adapt based on individual responses. Conversely, interviews provide deep, tailored insights but are time- and resource-intensive. AI-driven surveys offer an efficient, scalable solution that personalizes questions in real-time according to each respondent's answers, capturing richer, more nuanced data. This middle-ground approach addresses the need for both efficient data collection and depth of understanding, making it an appealing option for educational researchers who seek to balance scale with meaningful participant engagement.

To address \textbf{RQ1}, we present a step-by-step guide for integrating OpenAI’s ChatGPT API into the Qualtrics survey platform to generate personalized, contextually relevant questions. Our prompt engineering strategies are informed by educational theories---such as SSRL and effective team dynamics---to align question content with educational and research goals. We also outline key implementation considerations, including data security (e.g., encryption and authentication), error handling, and platform scalability. Ethical considerations are addressed through the introduction of the SQRA framework, which supports both design and validation of GenAI-based survey tools.

We apply the SQRA framework to evaluate survey performance prior to deployment with human participants. Using ``simulated participant rules,'' we generate synthetic student personas \emph{in silico}, enabling AI-to-AI interactions that serve as a testbed for iterative refinement. We then apply a range of analytical techniques---including sentiment analysis, cosine similarity, and structural analysis---to examine the quality, variation, and limitations of the AI-generated questions.

We use these analytical tools to compare the attributes of the AI-generated survey questions from AI-to-AI conversations to AI-to-Human conversations that were collected during the Fall 2024 semester from two courses. We use the theoretical lens of AT~\cite{Russell, DeCosta, Georg, ilishkina2025rethinking, Hite} to explore the bilateral relationship between our survey tool and the participant (either a human or AI agent). Through this analysis we observe both similarities and differences between AI-to-AI and AI-to-Human interactions. 

The sentiment analysis indicated that most responses in both AI-to-AI and AI-to-Human interactions were neutral, with positive outliers and minimal negative sentiment. We also found evidence that our generated questions successfully incorporated contextualization of our course activities. While the tool's neutrality and structured responses contribute to trustworthiness and consistency, they may lack the emotional depth or diversity seen in genuine human interactions. Furthermore, these limitations extend to the SQRA framework itself, where we see that the AI agents interact with the tool quite differently than the human participants.

We examined how well the generated questions aligned with the previous participant responses using cosine similarity scores. For both AI-to-AI and AI-to-Human interactions, we observed high cosine similarity scores, all above 0.5. However, human responses showed greater variability and, on average, lower similarity scores. Upon closer inspection, we found that brief responses---ranging from a single word to a sentence---accounted for all of the lower similarity outliers. This suggests that refining the ``simulated participant rules'' within the SQRA framework to occasionally emulate brief student responses could help better simulate realistic interactions. Brief reflective responses are well-documented in educational literature~\cite{franklin2018improving, may2002college, werth2023investigating, treibergs2022introductory, csavina2016assessing} and represent a key goal in developing this AI-generated survey instrument. This finding also has implications for the system prompt used to generate questions; for example, we could include clearer instructions on how to generate questions that effectively follow brief responses in an authentic way.

We analyzed question structure, focusing on character count and the frequency of certain symbols and conjunctions. We used conjunctions as a proxy for double-barreled questions, exclamation marks and periods as indicators of evaluative feedback, and multiple question marks as signs of multiple questions within one prompt---all of which were explicitly instructed to be avoided in our system prompt. On average, both AI-to-AI and AI-to-Human conversations contained about one question mark per question. However, AI-to-AI responses tended to be longer, with more frequent use of periods (approximately 0.5 per question) and the conjunction ``and'' (around 1.5 per question). This longer structure in AI-to-AI interactions appears to reflect the more elaborate responses generated by the simulated participant compared to human responses, which tend to be more concise.

AT provides a useful framework to interpret these limitations of the SQRA method, as the differing rules governing AI agents and human participants shape engagement differently. AI agents, constrained by system prompts and model architecture, tend to provide structured, expansive responses with minimal emotional variation, whereas human responses are influenced by social norms, cognitive biases, and interpersonal dynamics. Additionally, the community component of AT further contextualizes these findings, as AI agents lack the shared norms, hesitations, or implicit biases present in student responses. Unlike students, AI models do not navigate concerns such as uncertainty, self-reflection, or perceived social consequences of their responses. As a result, they may not exhibit the same skepticism or reluctance as human participants. This impacts the tool itself, as interactions with AI agents may reinforce certain question types that do not necessarily align with human cognitive and emotional variability.

To address this limitation, adjustments to model parameters---such as temperature settings or explicit prompting---may enhance tone variability, enabling AI to simulate more human-like emotional responses. These refinements ensure that AI-generated outputs align more closely with the intended use and audience, enhancing their role in stress-testing the survey. By applying AT as a guiding framework, we recognize that modifications to the participant alter the dynamics of the tool, reinforcing the need for iterative refinement based on both human participant and AI-driven interactions.

Throughout our analysis we were also able to identify several issues related to question quality that required further refinement. One recurring problem was the generation of double-barreled questions, which could lead to less reliable or interpretable data depending on the research objectives. Additionally, we identified redundancy in question phrasing, where AI-generated prompts often repeated similar structures and vocabulary across interactions. This repetition could potentially reduce student engagement and limit the depth of responses. Furthermore, AI-generated questions tended to be overly long and occasionally included implicit evaluations, such as affirmations or personal judgments. Another notable issue was the inclusion of jargon derived directly from the system prompt, such as the term ``positive interdependence.'' While the term was intended to encourage students to reflect on evidence-based principals of effective teamwork proposed in Adams (2002) work, its use in AI-generated questions was not ideal for a reflective prompt, as students unfamiliar with the terminology may have struggled to engage meaningfully with the question. 

To address the identified issues, we iteratively refined the system prompt, incorporating structured guidelines from Walther et al.'s (2017) quality framework for qualitative research. These modifications aimed to reduce redundancy, eliminate double-barreled questions, and remove implicit evaluative language. Through multiple rounds of refinement, we observed measurable improvements in question clarity, conciseness, and overall alignment with best practices in survey methodology. While the SQRA framework has inherent limitations in replicating human participant responses, our findings underscore the importance of system prompt refinement in AI-driven survey design. Moreover, we demonstrate the SQRA framework's utility as a method for critically evaluating and iterating on AI-generated questions to enhance their validity and interpretability.

\subsection{Implications and Methodological Contributions}

The development and implementation of AI-driven surveys in educational research carry several key implications and contributions, particularly as they relate to scalability, personalization, and need for new methodologies.

AI-driven surveys present a unique capability to bridge the gap between traditional surveys and interviews, balancing scale with personalization. Unlike conventional surveys, which lack adaptability to individual respondents, AI-driven surveys enable the dynamic generation of questions that respond to each participant's input, with the evidence-based goal of deepening participant engagement. This level of customization addresses a long-standing limitation in survey-based research and offers a novel method for educational researchers to capture nuanced insights without the extensive time and resources typical of interviews. The SQRA framework supports this scalability by allowing the pre-testing of AI-generated content, building trust and ensuring relevance prior to implementation in real-world settings.

The SQRA framework itself represents a significant methodological contribution. By incorporating simulated participant rules, sentiment analysis, cosine similarity, and question structure analysis, SQRA offers researchers a systematic approach to evaluating AI-generated survey instruments. This framework not only supports the generation of synthetic data that mirrors realistic student responses but also provides a way to iteratively refine system prompts. The framework's analytical tools allow researchers to examine alignment, tone, and structure, thus ensuring that AI-generated questions are both pedagogically aligned and contextually appropriate. SQRA could become an essential tool for researchers aiming to evaluate the effectiveness in AI-driven question generation.

Lastly, we applied the theoretical lens of AT to examine the dynamic interactions between the AI-driven survey tool and participants, whether AI agents or human respondents. AT provides a framework for understanding how various elements—such as the participant (subject), survey questions (tools), and the purpose of data collection (object)---interact within an educational context. While we mostly explored the interaction between the subject and tool, we believe that this theoretical lens opens numerous opportunities to explore the complex interplay between the GenAI survey design, respondent engagement, and socio-cultural factors such as rules, community, and divisions of labor. 

\subsection{Limitations and Future Work}

Despite the methodological contributions of this work, several limitations and opportunities for future exploration remain. First, we observed that synthetic responses generated within the SQRA framework lacked the depth and variability typically seen in human communication, with statistically significant differences and large effect sizes across nearly all analyses. To foster more nuanced and realistic synthetic responses, future work should focus on refining prompts to address the range of interactional nuances seen in AI-to-Human versus AI-to-AI interactions. Additionally, readability analyses, such as calculating Flesch Reading Ease and Flesch-Kincaid Grade Level scores, could offer a clearer view of the readability and accessibility of generated text~\cite{Mac2022ComparisonReadability}, helping to ensure that generated content aligns with educational accessibility standards. Moreover, exploring bias testing within generated questions and simulated personas, along with implementing more sophisticated emotional analysis techniques, could enhance the framework's capacity for assessing inclusivity and sentiment.

Furthermore, this study was limited by the capabilities of ChatGPT-3.5 for question generation, which, while robust, may not fully reflect the potential of more advanced models. Future research should explore the use of more advanced models, such as GPT-4o, GPT-4.1, GPT-4.5, or specialized domain-adapted LLMs, which may offer enhanced adaptability, tone modulation, and response relevance. Additionally, incorporating domain-specific knowledge bases could improve the contextual relevance of generated questions. Developing dynamic prompts that can self-adjust based on the tone and content of responses may further enhance the authenticity and engagement of AI-driven interactions.

Another limitation pertains to the absence of extensive field testing with diverse student populations. While the SQRA framework enables pre-deployment testing, real-world applications in varied educational contexts will provide additional insights into the AI-driven survey tool's effectiveness for producing higher quality data and reflections. As described in Figure~\ref{fig:figure1}, during the deployment of the survey we randomly assigned students to Group A, which engaged them with the AI-driven survey instrument, and Group B, predefined, static questions. In future studies, we plan to analyze the differential impact of the survey instrument on the responses from these two populations. 

AI-driven surveys fill a unique role in education research by combining the scalability of traditional surveys with the personalized engagement of interviews, but the potential challenges cannot be ignored. The lack of personal rapport in AI-driven surveys may lead to superficial responses, as these tools rely solely on prompt quality rather than relational trust-building. Additionally, poorly designed prompts risk introducing bias or failing to capture nuanced participant experiences. To address these limitations, future work could refine prompt validation methods, such as iterative testing, expert feedback, and frameworks like the SQRA. Combining AI-driven surveys with follow-up interviews could bridge the gap between efficiency and depth, using AI-generated data to inform targeted, in-depth exploration of emerging themes. These refinements will ensure that AI-driven surveys scale effectively while capturing the complexity and richness of human experiences, enhancing their reliability and trustworthiness in educational research.

Finally, while AT provided a valuable lens for analyzing interactional elements within the AI-driven survey framework, future studies could deepen this analysis by examining how each component of the activity system (subject, tool, object, community, rules, and division of labor) evolves dynamically in real-time responses. Notably, the inability to control all components of the activity system is not a limitation of this study but rather a reflection of the epistemological foundations underlying AT and qualitative research. These approaches embrace complexity and the dynamic, interconnected nature of systems, recognizing that full control or isolation of variables is neither feasible nor desirable for capturing the richness of real-world interactions. These approaches may offer richer insights into how students interpret and engage with AI-generated questions within a structured learning environment. Another promising avenue for future research is to explore how the AI-driven survey tool shapes participant behavior and, conversely, how participants adapt to the tool. AT provides a robust framework for examining this bidirectional influence, as it emphasizes the dynamic and co-evolving relationships within an activity system. Applying additional theoretical frameworks, such as naturalistic inquiry~\cite{LincolnGuba}, may also help explore how we establish trust and reliability in AI-driven survey methodologies.

These directions for future work will contribute to refining AI-driven survey instruments, enhancing their adaptability and utility across educational contexts. By addressing these limitations, we can more effectively integrate AI tools into educational research and practice, creating richer, more engaging, and educationally aligned survey experiences for diverse populations.

\section{Conclusion}

This study examined the integration of a LLM into Qualtrics to generate contextually relevant, personalized survey questions, focusing on scalability, security, and ethical use. Findings suggest that AI-driven surveys can bridge the gap between traditional surveys and interviews by combining the scale of survey data collection with the tailored engagement of interviews, with the goal of enhancing the depth and richness of data in educational research.

Using AT as a theoretical lens, we defined the core components of the activity system to explore the complex, dynamic interactions between subjects (student participants), tools (AI-driven surveys), and objects (participant responses). This approach emphasized the reciprocal influence between participants and the AI tool, illustrating how the tool mediates engagement while being refined by participant input. Future research will extend this framework to investigate the roles of community and division of labor, ensuring AT remains an actionable guide for analyzing and improving interactions in complex, tool-mediated environments. A primary contribution of this work is the development of the SQRA framework, which establishes a structured methodology for evaluating AI-generated survey questions prior to deployment. By employing sentiment analysis, cosine similarity, and structural analysis, we identified notable differences between AI-to-AI and AI-to-Human interactions. Although AI-generated questions showed promise in emulating human-like responses, variations in response depth and complexity underscore the need for ongoing refinement in prompt engineering.

In sum, this study emphasizes the expanding role of AI-driven surveys in educational research, advocating for methodological innovation to establish validity, reliability, and trustworthiness in AI outputs. These findings underscore the importance of balancing practical gains in scalability and adaptability with ethical considerations and the necessity for transparent, trustworthy AI-driven methods.

\section*{Acknowledgments}
This research was supported by the Center for Teaching Innovation (CTI) Innovation Grant for 2024-25 at Cornell University and the Cornell University Meinig School of Biomedical Engineering. We extend our gratitude to the students who participated in trialing this tool and contributed to the research. Special thanks to Natasha Holmes, René Kizilcec, Michelle Smith, and Kim Benowski for their contributions to this project and our research. We are also grateful to Noah Finkelstein, Angela Bielefeld, and Daniel Knight for their insightful guidance. Finally, we thank the Cornell Discipline-Based Education Research community for their ongoing advice and support and the Colorado Boulder Physics and Engineering Education Communities.

\section*{Financial disclosure}

None reported.

\section*{Conflict of interest}

The authors declare no potential conflict of interests.

\bibliographystyle{wileyNJD-APA}
\bibliography{wileyNJD-APA}

\clearpage
\section*{Appendix}
\addcontentsline{toc}{section}{Appendix}
\renewcommand{\thesection}{A\arabic{section}}
\setcounter{section}{0}

\section{Randomization and Treatment Groups}\label{app:randomization}
\begin{figure}[h]
\centering
\includegraphics[width=0.7\textwidth]{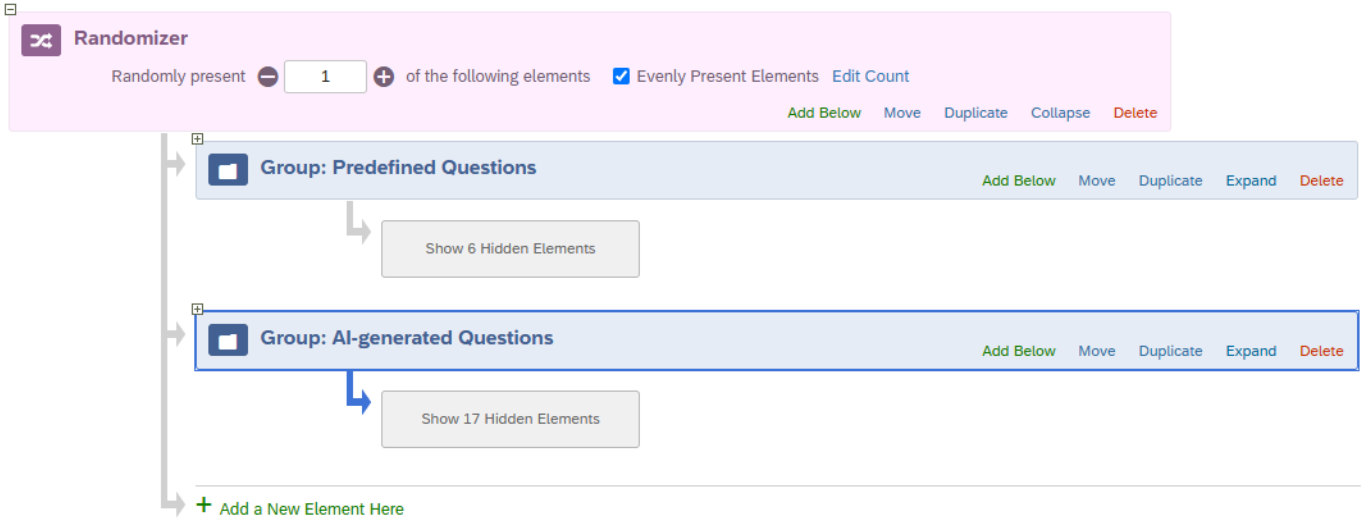}
\caption{Randomization process using Qualtrics.}
\label{fig:figure3Randomize}
\end{figure}

To minimize bias and ensure the validity of the study, participants were randomly assigned to either a control or a treatment group using Qualtrics' randomization feature. Randomization helps distribute potential confounding variables evenly across both groups, ensuring that any observed differences in responses are attributable to the intervention and not to participant characteristics. As demonstrated in  Figure \ref{fig:figure3Randomize}, by incorporating a Randomizer block within the survey flow, participants were randomly assigned to different groups, each exposed to different AI-generated questions. This method maintains the internal validity of the study by ensuring that each participant has an equal chance of receiving either the control or treatment condition.

The control group was exposed to standard, non-AI-generated questions, while the treatment group received dynamic, AI-generated questions tailored to their responses. This setup allowed us to assess the impact of AI-driven question generation compared to traditional methods. 

\begin{figure}[h]
\centering
\includegraphics[width=0.7\textwidth]{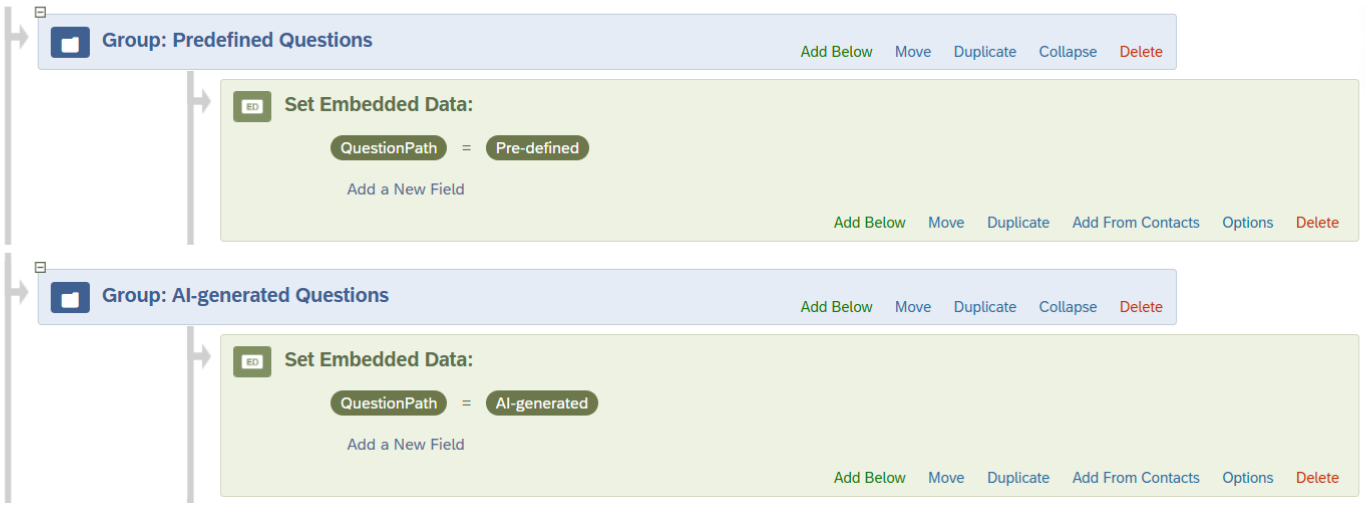}
\caption{Saving embedded data for each participant group type in Qualtrics. }
\label{fig:figure4}
\end{figure}

After the randomization, embedded data is used to track the participants' group assignments, allowing the researchers to filter and analyze responses based on group designation, see Figure \ref{fig:figure4}.

\newpage\section{Security and Privacy}\label{app:securityandprivacy} The OpenAI API used in this study to generate contextually relevant survey questions, also incorporates several advanced security measures to safeguard data. OpenAI has been audited for SOC 2 Type 2 compliance, which ensures that their systems meet security standards. Data encryption is implemented both at rest and in transit, using AES-256 for stored data and TLS 1.2+ for transmitted data. OpenAI does not train its models on API data by default unless explicitly opted-in by the user. Additionally, OpenAI's API offers a Data Processing Addendum (DPA) to support compliance with General Data Protection Regulation (GDPR) and other privacy laws. API inputs and outputs are securely retained for up to 30 days to identify potential abuse and maintain service integrity, after which they are permanently deleted unless required by law.

In this study, participant data was handled with strict adherence to privacy regulations, including the FERPA and the GDPR. All participant information was securely stored within Qualtrics, which provides robust security features for managing sensitive data. Qualtrics ensures compliance with these regulations by encrypting data both at rest and in transit using AES-256 encryption and TLS 1.2+. As seen in Figure \ref{fig:figure2}, Quatrics also offers secure user authentication through Single Sign-On (SSO) protocols, ensuring that participant data is only accessible by authorized personnel. No participant personal information like name, NetId, or email address, ever sent to the LLM to generate questions. Only deidentified survey responses are processed, protecting that data can not be traced back to individual participants.

\setcounter{figure}{0}
\renewcommand{\thefigure}{B\arabic{figure}}
\begin{figure}[h]
\centering
\includegraphics[width=0.6\textwidth]{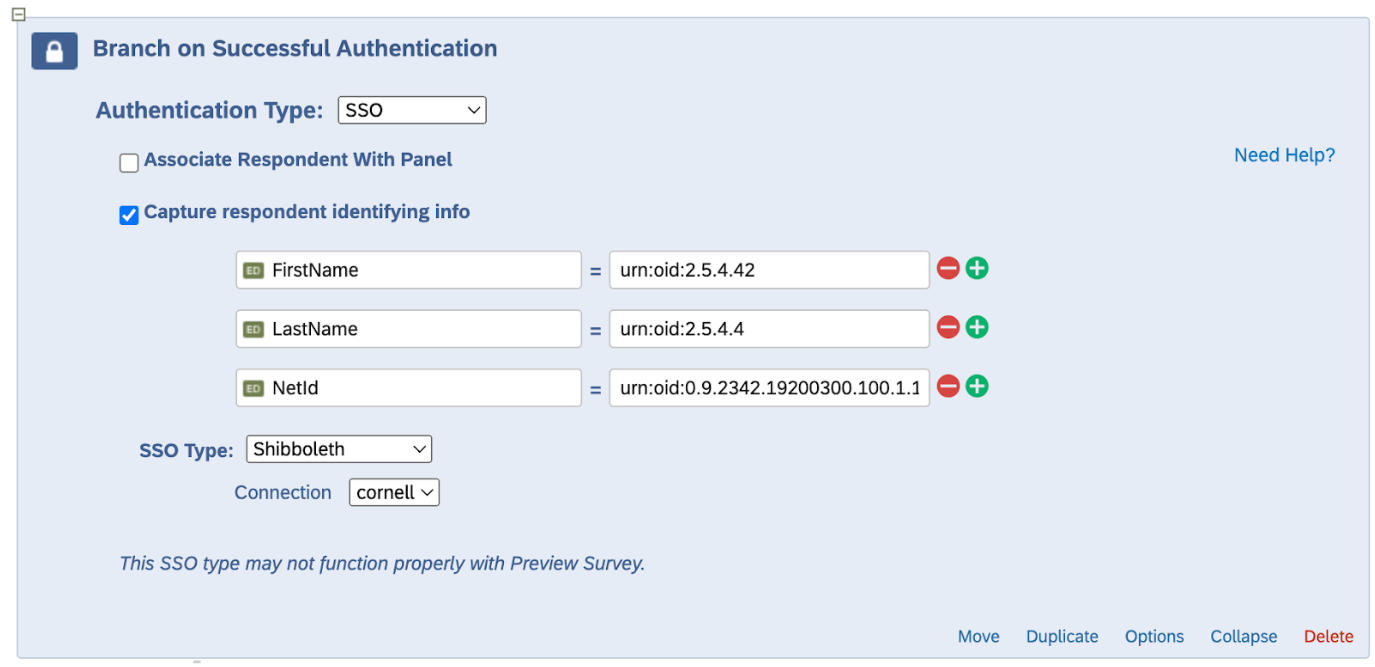}
\caption{Initial authenticator element to securely track student responses.}
\label{fig:figure2}
\end{figure}

\section{Embedded Data} \label{app:embeddeddata}
Embedded data played a crucial role in managing participant responses and streamlining the dynamic question-generation process within Qualtrics. These variables were set up to store key information, such as participant IDs, group assignments (control or treatment), and responses to individual survey questions. This structured data storage allowed for easy retrieval and reference throughout the survey flow, enabling personalized interactions for each participant. Embedded data could be predefined, like participant IDs or experimental conditions, or captured in real-time, such as survey responses or the time taken to complete a section. This allowed the system to dynamically tailor the survey content based on specific criteria, directing the flow through branch logic and managing the visibility of questions using display logic.

\section{Web Services}\label{app:webservices}

As demonstrated in Figure \ref{fig:figure5}, when using a web service, several key components define how data is exchanged with the LLM API. The URL specifies the endpoint of the API that the survey will interact with. This is different for other LLMs such as Gemini from Google or Claude from Anthropic. The `Method' is set to `POST,' indicating that the survey is sending data (such as user responses) to the API, as opposed to retrieving data. `Body Parameters' include specific details sent to the API, such as the model (e.g., GPT-3.5 or GPT-4) and messages, which contain the user's input and the system's instructions. Custom headers are additional pieces of information added to the request, such as authentication tokens, to ensure the API recognizes and securely processes the request. Finally, `Set Embedded Data' stores the API's response, whether it is a new question or an error allowing the data to be used dynamically in subsequent survey steps.

\setcounter{figure}{0}
\renewcommand{\thefigure}{D\arabic{figure}}
\begin{figure}[h]
\centering
\includegraphics[width=0.7\textwidth]{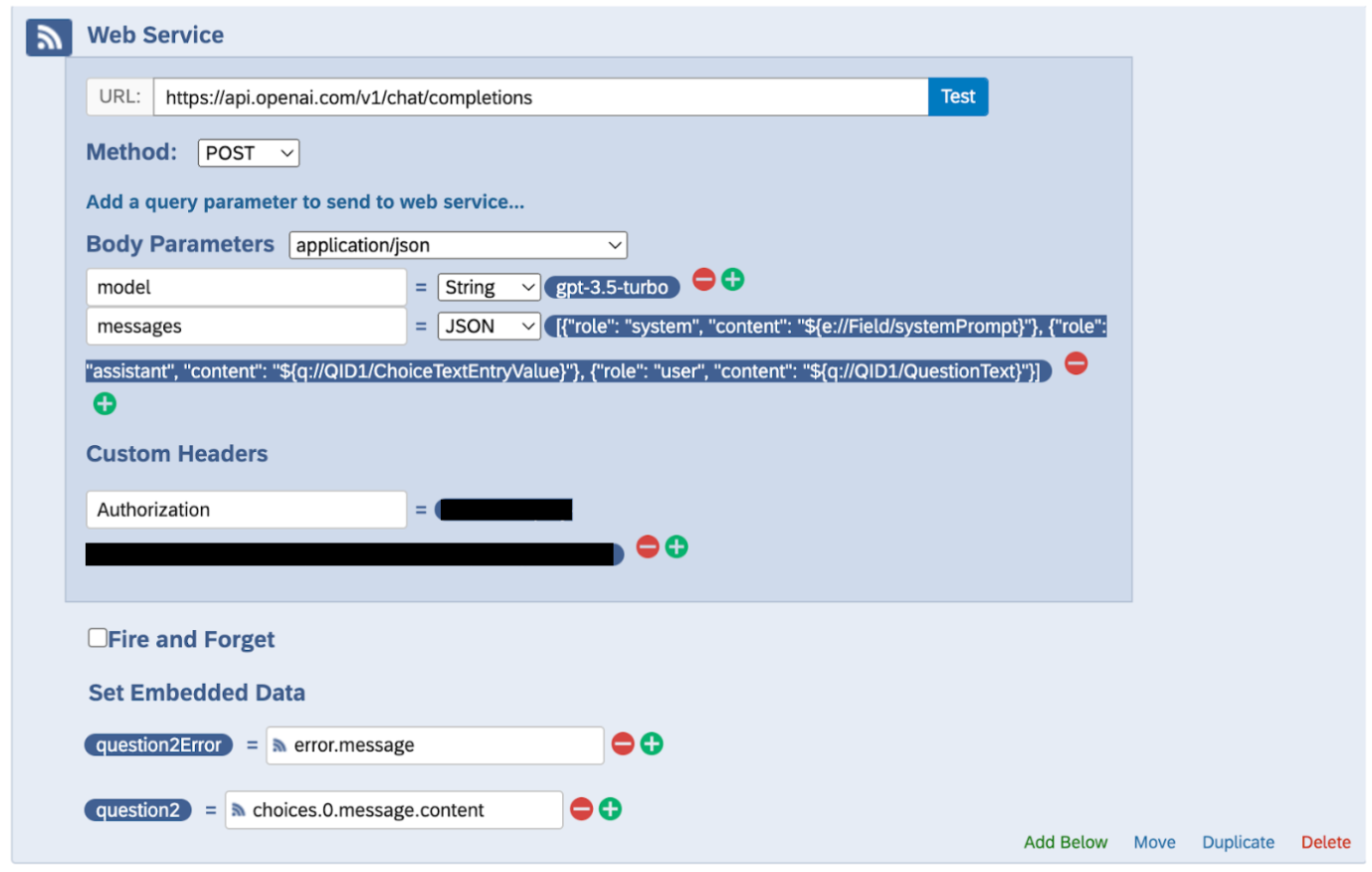}
\caption{Web service communication between Qualtrics and an external API.}
\label{fig:figure5}
\end{figure}

In Qualtrics, the `messages' body parameter structures the conversation history that is sent to the LLM API, enabling it to generate contextually relevant responses based on prior interactions. Each part of the parameter serves a purpose: (i) ``system'' provides the AI with instructions or context, typically using a system prompt stored as embedded data (\verb|$e://Field/systemPrompt|), (ii) ``assistant'' represents the AI's generated question or response, and (iii) the ``user'' captures the participant's input (e.g., {\verb|${q://QID1/ChoiceTextEntryValue|} for the first response).

\begin{figure}[h]
\centering
\includegraphics[width=0.7\textwidth]{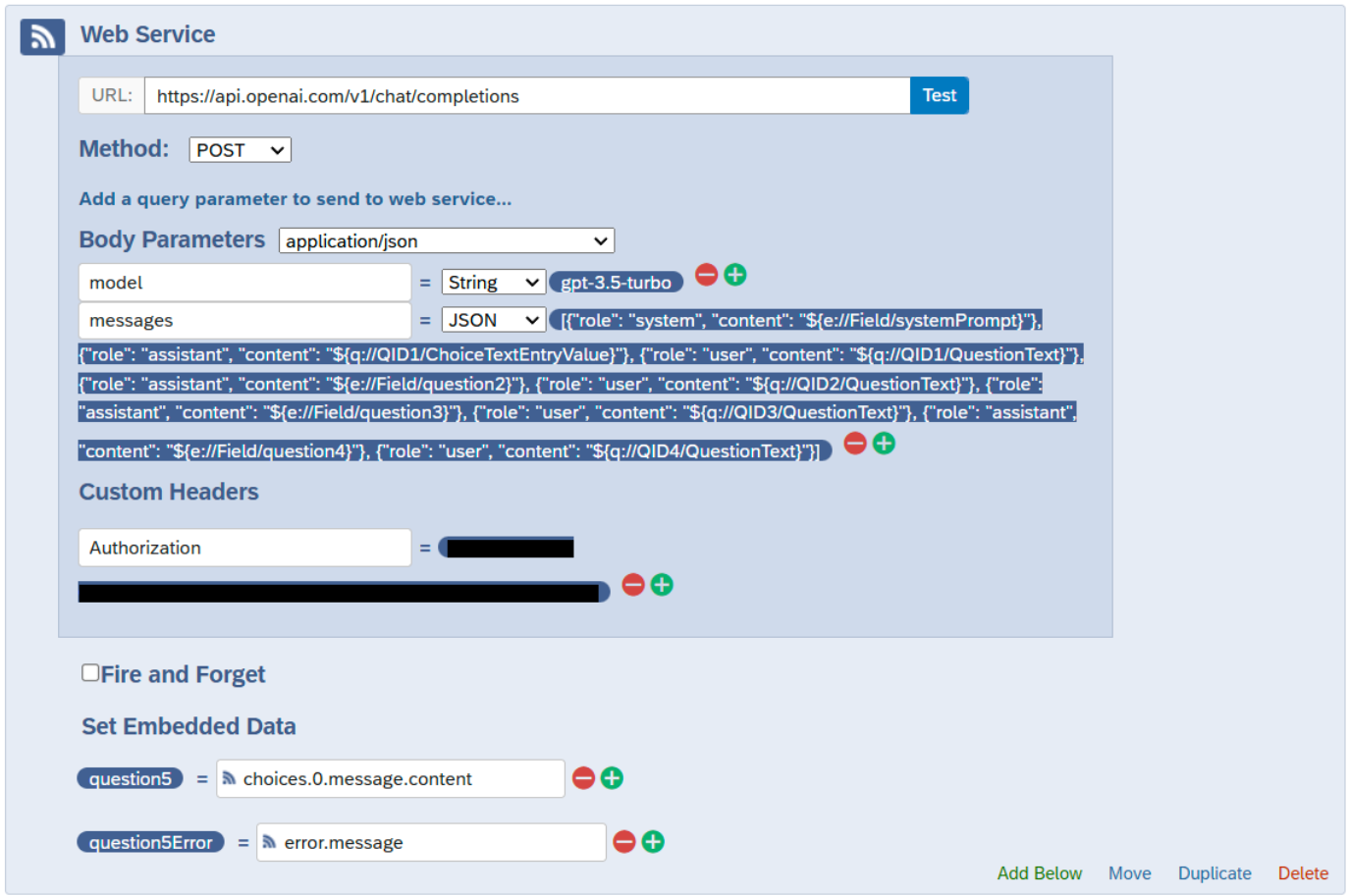}
\caption{``Messages'' body parameter changes between questions.}
\label{fig:figure6}
\end{figure}

For the first question, the messages parameter includes only the system prompt, the researcher's prediefined first question, and the participant's response. When transitioning to the second question, additional history is appended to the messages: the assistant's second question (\verb|${e://Field/question2}|) and the user's second response (\verb|${q://QID2/QuestionText}|). This process continues for each subsequent question, where each new AI-generated question and user response is added. For example, moving from the second to the third question, you append the assistant's response (\verb|${e://Field/question3}|) and the user's input (\verb|${q://QID3/QuestionText}|).  Figure \ref{fig:figure6} demonstrates what a Web Service for the fifth question in a survey would look like.

In this way, the conversation history grows with each new interaction, ensuring that the AI maintains the full context of the dialogue. This structured format allows the LLM to provide more coherent and contextually relevant responses by referencing all previous exchanges, making the conversation flow naturally and logically. This also means that each subsequent question requires more tokens than its predecessor.

\bmsubsection{Error Handling} \label{app:errorhandling}
In any research design involving external APIs, such as OpenAI's API for AI-driven question generation, there is always the risk of encountering issues such as timeouts, unavailability of the service, or unexpected responses. To ensure that these potential errors do not adversely affect participant experience or the integrity of the data collected, a fail-safe mechanism was implemented within the survey flow in Qualtrics.

This fail-safe mechanism detects any errors or slowdowns in API responses and automatically defaults to a set of pre-defined questions. This ensures that if the OpenAI API becomes unavailable or returns an error during a participant's interaction, the survey process continues smoothly without interruption,  preventing data loss and frustrated participants who may otherwise be stuck waiting for the system to respond. Pre-defined fallback questions, which are stored as embedded data within Qualtrics, are displayed to participants in place of the AI-generated questions (Figure \ref{fig:figure7}). Additionally, all API error messages and system responses were captured and saved within the survey's embedded data, providing us with a log of any issues encountered during data collection. 

\setcounter{figure}{0}
\renewcommand{\thefigure}{D\arabic{figure}}
\begin{figure}[h]
\centering
\includegraphics[width=0.7\textwidth]{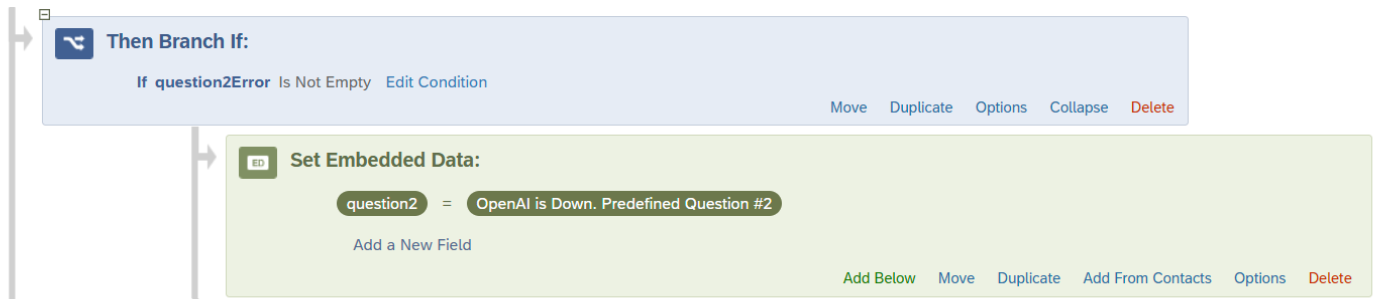}
\caption{Branch logic if OpenAI API is down from Embedded Data to a predefined question. }
\label{fig:figure7}
\end{figure}

\section{Demographic Data} \label{app:demographics}
Demographic data for this study was collected through separate surveys administered in the participating courses, rather than through the AI-driven questionnaire. Across both physics courses there was over a 90\% response rate on the demographic surveys and participation in the research related to the AI-generated survey. The biomedical engineering course had a 70\% response rate to both the demographic survey and participation in the research related to the AI-generated survey. This data can be found in Tables \ref{tab:demographics_summary} and \ref{tab:demographics}.

\begin{table}[h]
\centering
\caption{Demographic data for the biomedical engineering course.}
\begin{tabular}{|p{6cm}|c|}
\hline
\textbf{Category} & \textbf{n = 38} \\ 
\hline
\multicolumn{2}{|c|}{\textbf{Race/Ethnicity}} \\ \hline
Black or African American & 3 \\ 
East Asian & 12 \\ 
Hispanic, Latino, or Spanish & 6 \\ 
Middle Eastern or North African & 3 \\ 
South Asian & 4 \\ 
White & 10 \\ 
\hline
\multicolumn{2}{|c|}{\textbf{Education Level of Parent/Guardian}} \\ \hline
Bachelor's degree & 8 \\ 
High school diploma/GED & 4 \\ 
Less than a high school diploma & 1 \\ 
Master's degree or higher & 23 \\ 
Some college or an associate/trade degree & 2 \\ 
\hline
\multicolumn{2}{|c|}{\textbf{Gender}} \\ \hline
Man & 10 \\ 
Prefer not to answer & 1 \\ 
Woman & 27 \\ 
\hline
\multicolumn{2}{|c|}{\textbf{Years as a Student}} \\ \hline
3 & 36 \\ 
4 & 2 \\ 
\hline
\end{tabular}
\label{tab:demographics_summary}
\end{table}

\begin{table}[h]
\centering
\caption{Demographic data for the two physics courses.}
\begin{tabular}{|p{6cm}|c|c|}
\hline
\textbf{Category} & \textbf{n = 388} & \textbf{n = 40} \\ 
\hline
\multicolumn{3}{|c|}{\textbf{Race/Ethnicity}} \\ \hline
Black or African American & 12 & 1 \\ 
Asian & 219 & 19 \\ 
Hispanic or Latino & 24 & 5 \\ 
Middle Eastern or North African & 6 & 1 \\ 
White & 103 & 16 \\ 
\hline
\multicolumn{3}{|c|}{\textbf{Gender}} \\ \hline
Man & 175 & 24 \\ 
Woman & 157 & 8 \\ 
Nonbinary or Other & 8 & 2 \\ 
Did not disclose & 48 & 6 \\ 
\hline
\multicolumn{3}{|c|}{\textbf{Class Standing}} \\ \hline
Freshman & 279 & 1 \\ 
Sophomore & 66 & 34 \\ 
Junior & 7 & 2 \\ 
Senior & 2 & 1 \\ 
Did not disclose & 34 & 2 \\ 
\hline
\multicolumn{3}{|c|}{\textbf{Major}} \\ \hline
Physics, Astronomy, or Engineering Physics & 68 & 32 \\ 
Engineering & 227 & 3 \\ 
Life Science or Biology & 9 & 0 \\ 
Other Physical Science & 33 & 2 \\ 
Other & 5 & 0 \\ 
Did not disclose & 46 & 3 \\ 
\hline
\end{tabular}
\label{tab:demographics}
\end{table}

\section{Sentiment Analysis of AI-to-AI and AI-to-Human Interactions} \label{app:additionaldata}

To assess the sentiment distribution in AI-generated interactions, we performed sentiment analysis on both AI-to-AI and AI-to-Human exchanges. Figure \ref{fig:sentimentAnalysisPosNegNeut} illustrates the distribution of positive (green), neutral (blue), and negative (red) sentiments in each interaction type. Both types predominantly exhibited neutral sentiments, with a smaller proportion of positive sentiments and minimal negative sentiments.

\setcounter{figure}{0}
\renewcommand{\thefigure}{F\arabic{figure}}
\begin{figure}[h] \centering \includegraphics[width=0.8\textwidth]{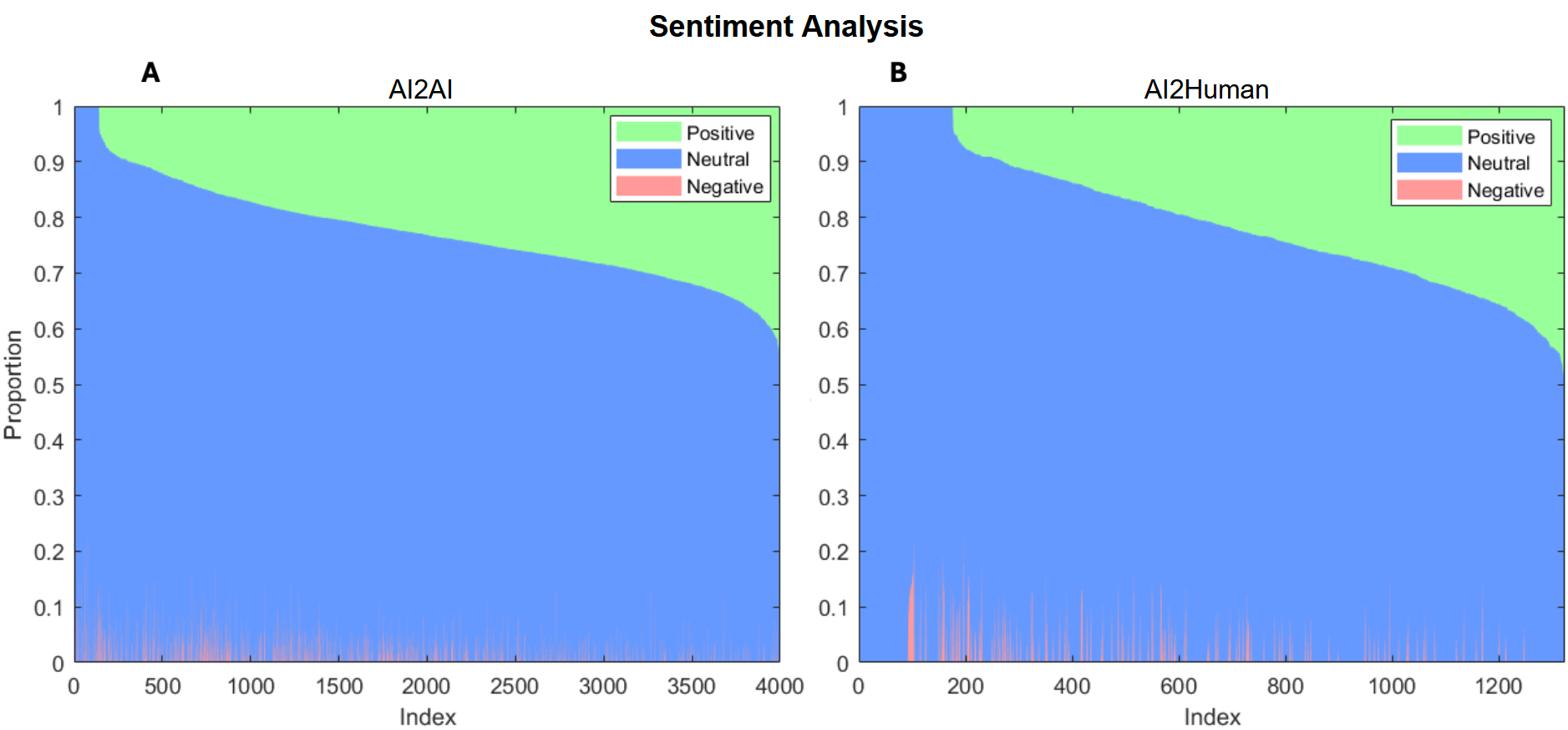} \caption{Sentiment analysis of AI-to-AI \textbf{A} and AI-to-Human \textbf{B} interactions. The distribution of positive, neutral, and negative sentiment scores across all responses is shown for each interaction type.} \label{fig:sentimentAnalysisPosNegNeut} \end{figure}

\section{Effect Sizes for Sentiment, Cosine Similarity, Symbol Frequency, and Character Count Comparisons}\label{app:effectsize} Table \ref{tab:Cohensd} presents the effect sizes (Cohen's $d$) for various aspects of the interaction analyses, allowing us to quantify differences across sentiment categories, cosine similarity of question-response pairs, symbol frequency, and character count. A positive effect size indicates a greater average value in Human-to-AI interactions, while a negative value indicates a greater average in AI-to-AI interactions.

\setcounter{table}{0}
\renewcommand{\thetable}{G\arabic{table}}
\begin{table}[h]
\centering
\caption{Effect Size (Cohen's $d$) for sentiment analysis, cosine similarity, symbol frequency, and character count comparisons of AI-to-AI and Human-to-AI populations.} 
\label{tab:Cohensd}
\begin{tabular}{|l|c|}
\hline
\textbf{Category} & \textbf{Effect Size (Cohen's $d$)} \\ 
\hline
\multicolumn{2}{|c|}{\textbf{Sentiment Analysis}} \\ \hline
Positive & 0.223 \\ 
Negative & 0.006 \\ 
Neutral & -0.227 \\ 
Compound & 0.640 \\ 
\hline
\multicolumn{2}{|c|}{\textbf{Response \& Question Similarity}} \\ \hline
R1 \& Q2 & 1.505 \\ 
R2 \& Q3 & 1.729 \\ 
R3 \& Q4 & 1.474 \\ 
R4 \& Q5 & 1.800 \\ 
\hline
\multicolumn{2}{|c|}{\textbf{Symbol Frequency}} \\ \hline
but & 0.113 \\ 
and & 1.547 \\ 
or & -0.132 \\ 
Exclamation mark & -0.110 \\ 
Period & 1.637 \\ 
Question mark & 0.317 \\ 
\hline
\multicolumn{2}{|c|}{\textbf{Character Count}} \\ \hline
Q2 & 2.700 \\ 
Q3 & 1.674 \\ 
Q4 & 1.423 \\ 
Q5 & 0.904 \\ 
\hline
\end{tabular}
\end{table}

\section{Outlier Texts} \label{app:cosinesimtext}

Table \ref{tab:characterCountQuestions} shows outliers from the character counts analysis. Tables \ref{tab:AI2HumancosineSimilarityOutliers} and \ref{tab:AI2AIcosineSimilarityOutliers} provide examples of outlier responses and associated follow-up questions, showing examples of the highest and lowest cosine similarity scores. 

\begin{table}[h!]
\centering
\caption{Outliers in sentiment analysis for AI-to-AI and AI-to-Human interactions.}
\label{tab:cosineOutliers}
\begin{tabular}{|p{2cm}|p{9cm}|p{1.4cm}|}
\hline
\textbf{Interaction Type} & \textbf{Question} &  \textbf{Cosine Similarity Score} \\ \hline
\vtop{\hbox{\strut AI-to-AI}\hbox{\strut (Most Negative)}} & How did openly discussing the workload distribution with your teammates impact your feelings of isolation and frustration within the team? & -0.70 \\ \hline
\vtop{\hbox{\strut AI-to-AI}\hbox{\strut (Most Positive)}} & That sounds like a thoughtful approach to improving teamwork dynamics. How do you think clearly defining roles and responsibilities can help us build stronger positive interdependence within our team? & 0.96 \\ \hline
AI-to-Human (Most Negative) & In reflecting on the team's process, how did your group handle disagreements or conflicting ideas when troubleshooting the code? & -0.59 \\ \hline
AI-to-Human (Most Positive) & How do you think each team member's strengths and roles contribute to the overall success of the team? & 0.75 \\ \hline
\end{tabular}
\end{table}

\clearpage

\setcounter{table}{0}
\renewcommand{\thetable}{H\arabic{table}}
\begin{table}[h!]
\centering
\caption{Outliers in character counts for questions in AI-to-AI and AI-to-Human interactions.}
\label{tab:characterCountQuestions}
\begin{tabular}{|p{1.7cm}|p{1.1cm}|p{10cm}|}
\hline
\textbf{Interaction Type} & \textbf{Character Count} & \textbf{Question} \\ \hline

AI-to-AI (Highest) & 609 & 
It's great that you've identified the need for clear expectations and a more inclusive environment within your team. Reflecting on your role and the dynamics at play is a crucial step towards fostering effective teamwork. In order to address the issue of dominant members and ensure everyone's voice is heard, have you thought about specific strategies or approaches that could help balance participation and encourage input from all team members? What are some ideas you have for establishing ground rules or facilitating discussions that could promote equal participation and collaboration within your team? \\ \hline

\vtop{\hbox{\strut AI-to-AI}\hbox{\strut (Lowest)}} & 85 & 
How have you tried to encourage participation and open communication within the team? \\ \hline

AI-to-Human (Highest) & 397 & 
It definitely made a positive impact on our teamwork this time. With a more cohesive approach, we were able to collaborate more effectively, share insights from our respective tasks, and troubleshoot together when issues arose. This led to a higher quality of work as we were able to combine our individual strengths and knowledge to produce a more thorough and well-rounded experiment and report. \\ \hline

AI-to-Human (Lowest) & 47 & 
How did the team adjust to having a new member? \\ \hline

\end{tabular}
\end{table}

\begin{table}[h]
\centering
\caption{Outliers in cosine similarity scores and associated responses in AI-to-Human interactions.}
\label{tab:AI2HumancosineSimilarityOutliers}
\begin{tabular}{|p{2cm}|p{1cm}|p{10cm}|}
\hline
\parbox{2cm}{\centering \textbf{Interaction \\ Type}} & 
\parbox{1cm}{\centering \textbf{Cosine \\ Similarity \\ Score}} & 
\parbox{6cm}{\centering \textbf{Previous Response and \\ Follow-up Question}} \\ \hline

\begin{tabular}[c]{@{}c@{}}AI-to-Human \\ (Highest)\end{tabular} & 
0.97 & 
\textbf{Previous Response:} To be fair, I know there is much to improve in my strategies to get them to refocus on the lab. I either worked through the report myself, or if I had questions I interrupted their conversation to ask for help. Sometimes I did have to ask multiple times before I got them to focus on my question. To their credit, they did offer great help once I did get them to refocus. Towards the last 30 minutes of our lab I started to be more assertive and push them to work since I got pressured by the time. I think by being more assertive, that helped them focus more on the lab. Ultimately however, they did not have the motivation to progress through the lab and I had to push them to complete it and be more ambitious with the experiments we were conducting.
\newline
\textbf{Follow-up Question:} How did you react when you had to push your teammates to work on the lab while you were typing up the report? How did that make you feel about your role within the team? \\ \hline

\begin{tabular}[c]{@{}c@{}}AI-to-Human \\ (Lowest)\end{tabular} & 
0.6054 & 
\textbf{Previous Response:} True \newline
\textbf{Follow-up Question:} That sounds like a positive change! How did being more responsive to each other's ideas impact the overall effectiveness of your team's work? \\ \hline

\end{tabular}
\end{table}

\begin{table}[t]
\centering
\caption{Outliers in cosine similarity scores and associated responses in AI-to-AI interactions.}
\label{tab:AI2AIcosineSimilarityOutliers}
\begin{tabular}{|p{1.5cm}|p{1cm}|p{10cm}|}
\hline
\parbox{1.5cm}{\centering \textbf{Interaction \\ Type}} & 
\parbox{1cm}{\centering \textbf{Cosine \\ Similarity \\ Score}} & 
\parbox{6cm}{\centering \textbf{Previous Response and \\ Follow-up Question}} \\ \hline

\begin{tabular}[c]{@{}c@{}}AI-to-AI \\ (Highest)\end{tabular} & 
0.99 & 
\textbf{Previous Response:} Reflecting on the team dynamics compared to last week, I notice a mix of progress and ongoing challenges. In terms of collaboration, there was a slight improvement as some members made an effort to involve everyone in discussions, allowing for more balanced participation. I appreciated this shift, as I felt more included in the brainstorming process and was able to share some of my ideas about our software solution. However, there were still moments when a few individuals dominated the conversation, leaving little room for others to contribute. The scheduling conflicts we faced last week continued to be an issue this week, making it difficult for everyone to be on the same page. This ongoing struggle has created an undercurrent of frustration, as it often feels like we are not fully capitalizing on everyone's strengths, including my own. While I sense some movement toward better teamwork, I still find myself navigating feelings of isolation and the burden of unequal workload distribution, which makes me even more determined to advocate for a more inclusive team dynamic moving forward. \newline
\textbf{Follow-up Question:} It's great to hear that there was some progress in including everyone in the discussions this week. It seems like you are feeling a mix of improvements and ongoing challenges in your team dynamics. Can you tell me more about how you plan to advocate for a more inclusive team dynamic moving forward, especially in addressing the concerns of domination in conversations and unequal workload distribution? \\ \hline

\begin{tabular}[c]{@{}c@{}}AI-to-AI \\ (Lowest)\end{tabular} & 
0.90 & 
\textbf{Previous Response:} When I approached my team to discuss the workload distribution and my feelings of being overlooked, I was honestly a bit nervous. I shared my perspective about how I often take on more responsibilities and highlighted the importance of everyone contributing equally for our project to succeed. To my surprise, they were receptive and acknowledged that they hadn't fully realized how it had affected me. We ended up having a constructive conversation where we clarified our individual roles and responsibilities moving forward, which helped ease some of the tension. It felt good to express my concerns honestly, even though I had to push through my anxiety about being assertive. It took some effort on everyone's part to agree on a schedule that accommodated our varying commitments, but I think this process really strengthened our communication. While I still sometimes feel like my contributions aren't fully recognized, this experience taught me the value of advocating for myself and the importance of asserting my voice in a team setting. Ultimately, it helped create a more collaborative environment where everyone understands each other's strengths and contributions better. \newline
\textbf{Follow-up Question:} How did clarifying individual roles and responsibilities help improve your team's communication and collaboration? \\ \hline

\end{tabular}
\end{table}

\end{document}